\documentclass[superscriptaddress,twocolumn,showpacs]{revtex4}
\usepackage{amsmath,amssymb}
\usepackage{graphicx}
\begin{document}
\title{Dynamical Properties of Quasi-One-Dimensional Boson-Fermion Mixtures of Atoms in a Toroidal Potential}
\author{Ryosuke Shibato}
\author{Takushi Nishimura}
\affiliation{Department of Physics, Tokyo Metropolitan University, Hachioji, Tokyo 192-0397, Japan}
\date{\today}
\begin{abstract}
We theoretically investigate quantum-mechanical dynamics of quasi-one-dimensional boson-fermion mixtures of atomic gases trapped in a toroidal potential, where effective inter-atomic interactions are tunable and affect the dynamics. We especially focus on effects of quantum statistics and many-body correlations beyond the Hartree-Fock (HF) mean-field approximation on the dynamics. In order to predict the dynamics, we utilize the numerical exact diagonalization method and also reproduce the calculation in the HF approximation for comparison. The toroidal gases originally have a rotational symmetry in the toroidal direction. We firstly prepare a deformed ground state as an initial state by adding a weak potential deformed in the toroidal direction, and then remove the potential to start the dynamics. In the dynamics, number densities of the deformed gases exhibit oscillations as demonstrated in the present paper. As a result, we find out that the bosons and fermions show quite different behaviors owing to quantum statistics. In particular, the bosons exhibit a low-frequency oscillation in the strong boson-boson attraction regime owing to the many-body correlations, and it can not be reproduced in the HF approximation. The oscillation of the fermions is strongly influenced by that of the bosons through the boson-fermion interaction as a forced oscillator. In addition, we also discuss a relationship between the low-frequency oscillation and restoration of the broken symmetry.
\end{abstract}
\maketitle
\renewcommand{\vec}[1]{\boldsymbol{#1}}
%
%
%
%
%
\section{Introduction \label{Sec-intro}}
According to quantum statistics, particles are classified into two types, i.e., bosons and fermions obeying the Bose-Einstein and Fermi-Dirac statistics, respectively. Bosons may occupy the same single-particle state and can produce the Bose-Einstein condensation, where macroscopic numbers of bosons occupy a single-particle state. On the other hand, fermions may not occupy the same single-particle state and can produce the Fermi degeneracy, where each fermion occupies a single-particle state up to the Fermi level. Quantum statistics plays an important role in quantum many-body physics not only for infinite matter systems but also for finite quantum systems. 

Recent developments of cooling and trapping techniques of atoms enable us to investigate quantum-statistical properties of ultra-cold atomic gases. In fact, atomic Bose-Einstein condensates~\cite{Bose}, Fermi degenerate gases~\cite{Fermi}, and boson-fermion mixtures~\cite{BFM_1, BFM_2, BFM_3, BFM_4, BFM_5, BFM_6, BFM_7, BFM_8, BFM_9, BFM_10} are realized and studied in actual experiments. 

As a feature of the ultra-cold atomic gases, precise optic and electromagnetic techniques have produced various quantum systems designed at will. In particular, the Feshbach resonance technique gives tunable effective inter-atomic interactions from attractions to repulsions~\cite{REV_1}. In addition, designed trap potentials have realized artificial crystal structures in the optical lattices and quasi-low-dimensional gases in the squeezed and toroidal potentials~\cite{REV_2}. 

Since bosons and fermions exhibit the quite different behaviors in the ultra-cold atomic gases, the boson-fermion mixtures present rich physics owing to the different quantum statistics and boson-fermion interactions. In fact, as static properties, rich phase structures are predicted in the atomic boson-fermion mixtures~\cite{BFM-the}. In addition, atomic fermions can produce the boson-fermion mixtures of atoms and molecules with quantum many-body correlations among atoms and molecules~\cite{nm, BF-pair}. Furthermore, as dynamical properties, collective excitations~\cite{sogo, mr} and collapse phenomena~\cite{miya, BFM_4} are studied theoretically and experimentally. 

In recent years, the ultra-cold atomic gases in the toroidal trap potential, called {\it toroidal} or {\it ring-shaped gases}, have been studied theoretically and experimentally. In fact, the toroidal bosons are experimentally demonstrated with the magnetic field~\cite{Mgn}, the optical field~\cite{Opt}, the magneto-optical field~\cite{MOT}, the microchip ring~\cite{chip}, and the ring-lattice~\cite{ring-lattice}. In addition, theoretical works on the toroidal bosons~\cite{Kanamoto, Toroidal_b} and boson-fermion mixtures~\cite{Riemann} are also published. 

As a feature of the toroidal gases, the gases can exhibit quasi-one-dimensional behaviors under a periodic boundary condition owing to geometry of the potential. Thus studies on the toroidal gases are closely connected to those on the one-dimensional systems. On another front, the toroidal gases are obviously a type of finite quantum many-body systems. In this sense, studies on the toroidal gases can also offer an important infrastructure for the finite quantum many-body problem. Our aim of the present study is near the latter side. 

In the quantum many-body problem, dynamical properties are fundamentally important. That is because the dynamics can give us significant information of quantum systems not only in theories but also in experiments. In fact, an indication of a spontaneous symmetry breaking (or phase transition) in the thermodynamical limit can be observed in the corresponding dynamics in finite quantum systems as demonstrated in the present paper and also, e.g., in nuclear physics~\cite{RS}. 

As an elementary knowledge of the quantum many-body physics, the mean-field approximation, which is conventionally used in studied on the cold atoms, gets worse near the phase transition in infinite matter systems and also near the corresponding situation in finite quantum systems. Furthermore, in parallel, the mean-field approximation also grows worse as the interactions get stronger. In fact, it is theoretically confirmed for the toroidal bosons~\cite{Kanamoto}. 

In the present paper, we theoretically study the dynamics of boson-fermion mixtures of the toroidal gases as a finite quantum many-body problem linked to the actual experiments. In the dynamics, we especially focus on effects of quantum statistics and many-body correlations beyond the mean-field approximation. 

In order to predict the dynamics, we utilize the numerical exact diagonalization method and also reproduce the calculation in the Hartree-Fock (HF) mean-field approximation for comparison. This procedure must be a conventional scheme to treat finite quantum systems and give exact solutions except for numeric errors and limitation of the configuration space on our computer memories. 

As explained above, the present study is closely connected to the theories of the one-dimensional systems. The theories are developed in mathematical physics~\cite{TG}. Especially those of the solvable models are studied well in the Bethe ansatz method~\cite{1D-The}, e.g., one-dimensional bosons~\cite{1D-The-2}, boson-fermion mixtures~\cite{1D-The-3}, and so on. In addition, those of the soliton solutions are also studied in the mean-field approximation~\cite{BF_soliton}. 

This paper is organized as follows. In Sec.~\ref{Sec-form}, we propose a theoretical formulation used in the present paper. In Sec.~\ref{Sec-static}, we show static properties of the toroidal gases as a background knowledge to study the dynamics. In Sec.~\ref{Sec-dynamic}, we demonstrate the dynamics in some typical situations. In Sec.~\ref{Sec-discussion}, we discuss the results obtained in the previous section. Section~\ref{Sec-summary} is devoted to summary and perspective. 

%
%
%
%
%
%
%
\section{Formulation \label{Sec-form}}
Here we formulate a theoretical model for the toroidal gases in Sec.~\ref{SubSec-TG} and its quantum-mechanical dynamics in Sec.~\ref{SubSec-D}, and also explain the calculational schemes used in the present paper in Sec.~\ref{SubSec-calculation}. In particular, the deformation parameters calculated in the next section are introduced in Sec.~\ref{SubSec-D}. 

In order to simplify the following descriptions, we determine the system of unit as $\hbar = 1$ for the reduced Planck constant without loss of generality.

\subsection{Toroidal gases \label{SubSec-TG}}
The toroidal gases are the dilute atomic gases trapped in the toroidal potentials in vacuum, where the typical size of the gases is much greater than that of the atomic scale, or the Bohr radius. 

In the cooling techniques of atoms, the gases are cooled down near zero temperature and exhibit purely quantum-mechanical behaviors. 

First we explain a theoretical treatment of the dilute atomic gases, and then derive the quasi-one-dimensional effective Hamiltonian in a conventional manner. 
\subsubsection{Theory of the dilute atomic gases \label{SubSec-TG-dag}}
The atoms of the dilute gases mostly behave as free particles except for extremely-short effective ranges of the inter-atomic interactions. In fact, the effective ranges are comparable with the atomic scale, and mean free paths of the atoms are much greater than the atomic scale in the dilute gases. 

In quantum mechanics, the many-body wave-function of the atoms exhibits singular behaviors near two-atom contact points owing to the extremely-short-range interactions, and the singularities themselves are mostly independent of gaseous behaviors because of the great difference of the length scales. The singularities mostly agree with those in the inter-atomic scattering problem. 

In order to simplify descriptions for the gaseous behaviors, we consider an asymptotic wave-function, instead of the real many-body wave-function, in the following manner. First we theoretically assume the asymptotic wave-function derived from the real wave-function by getting rid of the singularities. Second we renormalize the inter-atomic interactions with the corresponding low-energy scattering data, where information of the singularities are effectively included in the scattering data. 

The asymptotic wave-function approximately agrees with the wave-function of the non-interacting particles except for non-singular effects of the renormalized inter-atomic interactions. As a result, we can utilize the perturbation theory and mean-field approximation for the asymptotic wave-function unlike the case of the real wave-function. 

In the notations of the second quantization, we use the asymptotic field operators corresponding to the asymptotic wave-function, instead of the real field operators, in a conventional manner~\cite{FW}. 
\subsubsection{The quasi-one-dimensional effective Hamiltonian \label{SubSec-TG-H}}
Because the asymptotic fields roughly agree with those of the non-interacting particles as explained above, we first consider the toroidal gases of the non-interacting particles, and then introduce the renormalized effective interactions in a conventional manner. 

Let us consider the mixed gases of the single-component bosons and fermions in the toroidal potentials denoted by 
\begin{equation}
v_{\alpha}(\vec{r}) 
\equiv v^{(c)}_{\alpha}(r_{c}) + v^{(z)}_{\alpha}(r_{z}) 
\label{Eq-TG-H-1}
\end{equation}
for $\alpha = \text{b}$ and $\text{f}$, where the subscripts $\text{b}$ and $\text{f}$ indicate the potentials for the bosons and fermions, respectively. The potentials $v_{\alpha}(\vec{r})$ in Eq.~(\ref{Eq-TG-H-1}) are separated into the transverse and longitudinal potentials, $v^{(c)}_{\alpha}(r_{c})$ and $v^{(z)}_{\alpha}(r_{z})$, in the cylindrical coordinates, $\vec{r} = (r_{c}, \theta, r_{z})$, and independent of the azimuthal angle $\theta$. 

The transverse potential $v^{(c)}_{\alpha}(r_{c})$ in Eq.~(\ref{Eq-TG-H-1}) has a deep valley centering at $r_{c} = R$ with the toroidal radius $R$, and can be described as 
\begin{equation}
v^{(c)}_{\alpha}(r_{c}) 
\approx \frac{1}{2} m_{\alpha} {\omega^{(c)}_{\alpha}}^{2} \left( r_{c} - R \right)^{2} 
\label{Eq-TG-H-2}
\end{equation}
with the transverse trap frequencies $\omega^{(c)}_{\text{b}}$ and $\omega^{(c)}_{\text{f}}$ near the center of the valley. Here we introduce the boson and fermion masses $m_{\text{b}}$ and $m_{\text{f}}$, respectively. Since the gases are mostly trapped in the valley, the detailed profiles of the transverse potentials hardly affect the following formulation.

The longitudinal potential $v^{(z)}_{\alpha}(r_{z})$ in Eq.~(\ref{Eq-TG-H-1}) can be denoted by 
\begin{equation}
v^{(z)}_{\alpha}(r_{z}) 
= \frac{1}{2} m_{\alpha} {\omega^{(z)}_{\alpha}}^{2} r_{z}^{2} 
\label{Eq-TG-H-3}
\end{equation}
with the longitudinal trap frequencies $\omega^{(z)}_{\text{b}}$ and $\omega^{(z)}_{\text{f}}$. 

The Hamiltonian for the non-interacting bosons and fermions is denoted by 
\begin{equation}
H_{0} 
= H_{\text{b}} + H_{\text{f}} 
\label{Eq-TG-H-4}
\end{equation}
with the boson and fermion parts, 
\begin{equation}
H_{\alpha} 
= \int d{\vec{r}}~ \psi_{\alpha}^{\dagger}(\vec{r}) h_{\alpha}(\vec{r}) \psi_{\alpha}(\vec{r}) 
\label{Eq-TG-H-5}
\end{equation}
for $\alpha = \text{b}$ and $\text{f}$, where we introduce the single-particle Hamiltonian 
\begin{equation}
h_{\alpha}(\vec{r}) 
\equiv - \frac{1}{2 m_{\alpha}} \left( \frac{1}{r_{c}} \frac{\partial}{\partial{r_{c}}} r_{c} \frac{\partial}{\partial{r_{c}}} + \frac{1}{r_{c}^{2}} \frac{\partial^{2}}{\partial{\theta}^{2}} + \frac{\partial^{2}}{\partial{r_{z}}^{2}} \right) + v_{\alpha}(\vec{r}). 
\label{Eq-TG-H-6}
\end{equation}

The asymptotic field operators $\psi_{\text{b}}(\vec{r})$ and $\psi_{\text{f}}(\vec{r})$ for the bosons and fermions, respectively, obey the commutation and anti-commutation relations, 
\begin{equation}
\big[ \psi_{\text{b}}(\vec{r}), \psi_{\text{b}}(\vec{s}) \big]_{-} 
= 0,
~~ 
\big[ \psi_{\text{b}}(\vec{r}), \psi_{\text{b}}^{\dagger}(\vec{s}) \big]_{-} 
= \delta{(\vec{r} - \vec{s})}, 
\label{Eq-TG-H-7}
\end{equation}
\begin{equation}
\big[ \psi_{\text{f}}(\vec{r}), \psi_{\text{f}}(\vec{s}) \big]_{+} 
= 0,
~~ 
\big[ \psi_{\text{f}}(\vec{r}), \psi_{\text{f}}^{\dagger}(\vec{s}) \big]_{+} 
= \delta{(\vec{r} - \vec{s})}, 
\label{Eq-TG-H-8}
\end{equation}
and 
\begin{equation}
\big[ \psi_{\text{b}}(\vec{r}), \psi_{\text{f}}(\vec{s}) \big]_{-} 
= 0,
~~ 
\big[ \psi_{\text{b}}(\vec{r}), \psi_{\text{f}}^{\dagger}(\vec{s}) \big]_{-} 
= 0. 
\label{Eq-TG-H-9}
\end{equation}

The eigen-states of the single-particle Hamiltonian $h_{\alpha}(\vec{r})$ in Eq.~(\ref{Eq-TG-H-6}) can be described as 
\begin{equation}
\left[ h_{\alpha}(\vec{r}) - \varepsilon_{\alpha n m \kappa} \right] \phi^{(c)}_{\alpha n \kappa}(r_{c}) \phi^{(z)}_{\alpha m}(r_{z}) \phi^{(\theta)}_{\kappa}(\theta) 
= 0 
\label{Eq-TG-H-10}
\end{equation}
with the eigen-energy $\varepsilon_{\alpha n m \kappa}$, radial part $\phi^{(c)}_{\alpha n \kappa}(r_{c})$, longitudinal part $\phi^{(z)}_{\alpha m}(r_{z})$ and azimuthal part 
\begin{equation}
\phi^{(\theta)}_{\kappa}(\theta) 
\equiv \frac{e^{i \kappa \theta}}{\sqrt{2 \pi}} 
\label{Eq-TG-H-11}
\end{equation}
for the quantum numbers $n$ ($\ge 0$), $m$ ($\ge 0$), and $\kappa$. These parts are orthonormal as 
\begin{equation}
\int_{0}^{\infty} d{r_{c}} r_{c} \left[ \phi^{(c)}_{\alpha n \kappa}(r_{c}) \right]^{*} \phi^{(c)}_{\alpha n^{\prime} \kappa}(r_{c}) 
= \delta_{n, n^{\prime}}, 
\label{Eq-TG-H-12}
\end{equation}
\begin{equation}
\int_{- \infty}^{\infty} d{r_{z}} \left[ \phi^{(z)}_{\alpha m}(r_{z}) \right]^{*} \phi^{(z)}_{\alpha m^{\prime}}(r_{z}) 
= \delta_{m, m^{\prime}}, 
\label{Eq-TG-H-13}
\end{equation}
and 
\begin{equation}
\int_{0}^{2 \pi} d{\theta} \left[ \phi^{(\theta)}_{\kappa}(\theta) \right]^{*} \phi^{(\theta)}_{\kappa^{\prime}}(\theta) 
= \delta_{\kappa, \kappa^{\prime}}. 
\label{Eq-TG-H-14}
\end{equation}

By using complete sets of the eigen-states in Eq.~(\ref{Eq-TG-H-10}), we expand the field operators as 
\begin{equation}
\psi_{\alpha}(\vec{r}) 
= \psi_{\alpha}^{(P)}(\vec{r}) + \psi_{\alpha}^{(Q)}(\vec{r}) 
\label{Eq-TG-H-15}
\end{equation}
with the toroidal part 
\begin{equation}
\psi_{\alpha}^{(P)}(\vec{r}) 
\equiv \sum_{\kappa = - \infty}^{\infty} \phi^{(c)}_{\alpha 0 \kappa}(r_{c}) \phi^{(z)}_{\alpha 0}(r_{z}) \phi^{(\theta)}_{\kappa}(\theta) a_{\alpha 0 0 \kappa} 
\label{Eq-TG-H-16}
\end{equation}
and residual part
\begin{equation}
\psi_{\alpha}^{(Q)}(\vec{r}) 
\equiv \sum_{n = 1}^{\infty} \sum_{m = 1}^{\infty} \sum_{\kappa = - \infty}^{\infty} \phi^{(c)}_{\alpha n \kappa}(r_{c}) \phi^{(z)}_{\alpha m}(r_{z}) \phi^{(\theta)}_{\kappa}(\theta) a_{\alpha n m \kappa}. 
\label{Eq-TG-H-17}
\end{equation}
According to Eqs.~(\ref{Eq-TG-H-7})-(\ref{Eq-TG-H-9}), the particle operators, $a_{\text{b} n m \kappa}$ and $a_{\text{f} n m \kappa}$, obey the commutation and anti-commutation relations, $\big[ a_{\text{b} n m \kappa}, a_{\text{b} n^{\prime} m^{\prime} \kappa^{\prime}}^{\dagger} \big]_{-} = \delta_{n, n^{\prime}} \delta_{m, m^{\prime}} \delta_{\kappa, \kappa^{\prime}}$ and so on.

In the toroidal gases, although the interactions break the variable separation form in Eq.~(\ref{Eq-TG-H-10}), the particles mostly occupy the $n = m = 0$ states in the asymptotic field picture. That is because the excitation energies, $\varepsilon_{\alpha 1 0 \kappa} - \varepsilon_{\alpha 0 0 \kappa}$ and $\varepsilon_{\alpha 0 1 \kappa} - \varepsilon_{\alpha 0 0 \kappa}$, are much greater than both of the Fermi energy and absolute values of the interaction energies. 

At that time, we can focus on the limited configuration space described by $\psi_{\text{b}}^{(P)}(\vec{r})$ and $\psi_{\text{f}}^{(P)}(\vec{r})$, instead of the full configuration space, as an effective theory. This theory can be reduced to a type of the effective interaction theory~\cite{TNTM}. 

The toroidal parts $\psi_{\text{b}}^{(P)}(\vec{r})$ and $\psi_{\text{f}}^{(P)}(\vec{r})$ in Eq.~(\ref{Eq-TG-H-16}) can be rewritten as 
\begin{equation}
\psi_{\text{b}}^{(P)}(\vec{r}) 
= \phi^{(c)}_{\text{b} 0 0}(r_{c}) \phi^{(z)}_{\text{b} 0}(r_{z}) \sum_{\kappa = - \infty}^{\infty} \phi^{(\theta)}_{\kappa}(\theta) b_{\kappa} + O{(R^{-3})} 
\label{Eq-TG-H-18}
\end{equation}
and 
\begin{equation}
\psi_{\text{f}}^{(P)}(\vec{r}) 
= \phi^{(c)}_{\text{f} 0 0}(r_{c}) \phi^{(z)}_{\text{f} 0}(r_{z}) \sum_{\kappa = - \infty}^{\infty} \phi^{(\theta)}_{\kappa}(\theta) c_{\kappa} + O{(R^{-3})} 
\label{Eq-TG-H-19}
\end{equation}
with $b_{\kappa} \equiv a_{\text{b} 0 0 \kappa}$ and $c_{\kappa} \equiv a_{\text{f} 0 0 \kappa}$. Here 
\begin{equation}
\phi^{(c)}_{\alpha 0 0}(r_{c}) 
\approx \left( \frac{m_{\alpha} \omega^{(c)}_{\alpha}}{\pi R^{2}} \right)^{1 / 4} e^{- m_{\alpha} \omega^{(c)}_{\alpha} (r_{c} - R)^{2} / 2}, 
\label{Eq-TG-H-20}
\end{equation}
and 
\begin{equation}
\phi^{(z)}_{\alpha 0}(r_{z}) 
= \left( \frac{m_{\alpha} \omega^{(z)}_{\alpha}}{\pi} \right)^{1 / 4} e^{- m_{\alpha} \omega^{(z)}_{\alpha} r_{z}^{2} / 2}. 
\label{Eq-TG-H-21}
\end{equation}

In the limited configuration space, the $H_{\text{b}}$ and $H_{\text{f}}$ in Eq.~(\ref{Eq-TG-H-5}) are effectively reduced to 
\begin{equation}
H_{\text{b}} 
\approx \sum_{\kappa = - \infty}^{\infty} \frac{\kappa^{2}}{2 m_{\text{b}} R^{2}} b_{\kappa}^{\dagger} b_{\kappa} + O{(R^{-3})} + \mathcal{E}_{\text{b}} 
\label{Eq-TG-H-22}
\end{equation}
and 
\begin{equation}
H_{\text{f}} 
\approx \sum_{\kappa = - \infty}^{\infty} \frac{\kappa^{2}}{2 m_{\text{f}} R^{2}} c_{\kappa}^{\dagger} c_{\kappa} + O{(R^{-3})} + \mathcal{E}_{\text{f}} 
\label{Eq-TG-H-23}
\end{equation}
with the constant energies $\mathcal{E}_{\text{b}}$ and $\mathcal{E}_{\text{f}}$. The constant energies do not affect physics in the nonrelativistic systems. 

Because the higher terms of the $R$ inverse, $O{(R^{-3})}$, must have negligible contributions to physics of the toroidal gases, we neglect them in the following formulation. At the same time, the quantum number $\kappa$ should be limited in $\left| \kappa \right| \le \kappa_{c}$ with a large cut-off number $\kappa_{c}$ in order to justify the approximations in Eqs.~(\ref{Eq-TG-H-18}) and (\ref{Eq-TG-H-19}). 

As a result, the effective Hamiltonians in Eqs.~(\ref{Eq-TG-H-22}) and (\ref{Eq-TG-H-23}) indicate those of the quasi-one-dimensional gases of the non-interacting bosons and fermions. 

Next we consider the renormalized effective interactions $H_{\text{bb}}$ and $H_{\text{bf}}$ in the limited configuration space, where $H_{\text{bb}}$ and $H_{\text{bf}}$ indicate the boson-boson and boson-fermion interactions, respectively. Note that the interaction between the fermions is neglected owing to the Pauli blocking effect.

In a conventional manner, we utilize the renormalized contact-type pseudo-potentials for the low-energy $s$-wave scatterings, i.e., 
\begin{equation}
H_{\text{bb}} 
= \frac{\tilde{g}_{\text{bb}}}{2} \int d{\vec{r}}~ {\psi_{\text{b}}^{(P)}}^{\dagger}(\vec{r}) {\psi_{\text{b}}^{(P)}}^{\dagger}(\vec{r}) \psi_{\text{b}}^{(P)}(\vec{r}) \psi_{\text{b}}^{(P)}(\vec{r}) 
\label{Eq-TG-H-24}
\end{equation}
and 
\begin{equation}
H_{\text{bf}} 
= \tilde{g}_{\text{bf}} \int d{\vec{r}}~ {\psi_{\text{b}}^{(P)}}^{\dagger}(\vec{r}) {\psi_{\text{f}}^{(P)}}^{\dagger}(\vec{r}) \psi_{\text{f}}^{(P)}(\vec{r}) \psi_{\text{b}}^{(P)}(\vec{r}). 
\label{Eq-TG-H-25}
\end{equation}

The renormalized coupling constants $\tilde{g}_{\text{bb}}$ and $\tilde{g}_{\text{bf}}$ in Eqs.~(\ref{Eq-TG-H-24}) and (\ref{Eq-TG-H-25}) are determined as 
\begin{equation}
\tilde{g}_{\text{bb}} 
= \frac{4 \pi}{m_{\text{b}}} a_{\text{bb}} 
\label{Eq-TG-H-26}
\end{equation}
and 
\begin{equation}
\tilde{g}_{\text{bf}} 
= \frac{2 \pi (m_{\text{b}} + m_{\text{f}})}{m_{\text{b}} m_{\text{f}}} a_{\text{bf}} 
\label{Eq-TG-H-27}
\end{equation}
with the boson-boson and boson-fermion $s$-wave scattering lengths $a_{\text{bb}}$ and $a_{\text{bf}}$, respectively. The proportionality relations in Eqs.~(\ref{Eq-TG-H-26}) and (\ref{Eq-TG-H-27}) must be valid for the weakly-interacting gases as confirmed in the perturbation theory and mean-field approximation, or the Bethe-Goldstone theory, and should be verified for the strongly-interacting gases experimentally. 

It should be noted that the contact-type interactions in Eqs.~(\ref{Eq-TG-H-24}) and (\ref{Eq-TG-H-25}) can not produce divergence in the two-body correlations because of the limited configuration space unlike the case of the full configuration space. The divergence in the full configuration space originally comes from the singularities owing to the extremely-short-range interactions and must be removed in the asymptotic field picture. 

The interaction parts $H_{\text{bb}}$ and $H_{\text{bf}}$ in Eqs.~(\ref{Eq-TG-H-24}) and (\ref{Eq-TG-H-25}) are reduced to 
\begin{equation}
H_{\text{bb}} 
\approx \frac{g_{\text{bb}}}{4 \pi R} \sum_{\kappa_{1}} \sum_{\kappa_{2}} \sum_{\kappa_{3}} \sum_{\kappa_{4}} \delta_{(\kappa_{1} + \kappa_{2}), (\kappa_{3} + \kappa_{4})} b_{\kappa_{1}}^{\dagger} b_{\kappa_{2}}^{\dagger} b_{\kappa_{3}} b_{\kappa_{4}} 
\label{Eq-TG-H-28}
\end{equation}
and 
\begin{equation}
H_{\text{bf}} 
\approx \frac{g_{\text{bf}}}{2 \pi R} \sum_{\kappa_{1}} \sum_{\kappa_{2}} \sum_{\kappa_{3}} \sum_{\kappa_{4}} \delta_{(\kappa_{1} + \kappa_{2}), (\kappa_{3} + \kappa_{4})} b_{\kappa_{1}}^{\dagger} c_{\kappa_{2}}^{\dagger} c_{\kappa_{3}} b_{\kappa_{4}} 
\label{Eq-TG-H-29}
\end{equation}
with the redefined coupling constants 
\begin{equation}
g_{\text{bb}} 
\approx \frac{m_{\text{b}} \sqrt{\omega^{(c)}_{\text{b}} \omega^{(z)}_{\text{b}}} \tilde{g}_{\text{bb}}}{2 \pi} 
\label{Eq-TG-H-30}
\end{equation}
and 
\begin{equation}
g_{\text{bf}} 
\approx \frac{m_{\text{b}} m_{\text{f}} \sqrt{\omega^{(c)}_{\text{b}} \omega^{(z)}_{\text{b}} \omega^{(c)}_{\text{f}} \omega^{(z)}_{\text{f}}} \tilde{g}_{\text{bf}}}{\pi \sqrt{m_{\text{b}} \omega^{(c)}_{\text{b}} + m_{\text{f}} \omega^{(c)}_{\text{f}}} \sqrt{m_{\text{b}} \omega^{(z)}_{\text{b}} + m_{\text{f}} \omega^{(z)}_{\text{f}}}} 
\label{Eq-TG-H-31}
\end{equation}
according to Eqs.~(\ref{Eq-TG-H-18})-(\ref{Eq-TG-H-21}). 

Finally we obtain the quasi-one-dimensional effective Hamiltonian denoted by 
\begin{equation}
H 
= H_{\text{b}} + H_{\text{f}} + H_{\text{bb}} + H_{\text{bf}}, 
\label{Eq-TG-H-32}
\end{equation}
where these parts are obtained in Eqs.~(\ref{Eq-TG-H-22}), (\ref{Eq-TG-H-23}), (\ref{Eq-TG-H-28}) and (\ref{Eq-TG-H-29}). 

\subsection{Quantum-mechanical dynamics \label{SubSec-D}}

In quantum mechanics, dynamics is described as an initial value problem for the Schr\"{o}dinger equation, 
\begin{equation}
i \frac{d}{d{t}} \big| \Psi(t) \big> 
= H \big| \Psi(t) \big> 
\label{Eq-D-1}
\end{equation}
with the time-dependent state $\big| \Psi(t) \big>$. Here the Hamiltonian $H$ is given in Eq.~(\ref{Eq-TG-H-32}) in the present paper. 

In order to obtain the state $\big| \Psi(t) \big>$ in Eq.~(\ref{Eq-D-1}), we need to give an initial state $\big| \Psi(t_{0}) \big>$ at the initial time $t_{0}$. In general, quantum-mechanical dynamics depends on the initial state. 

Although we can freely give the initial state in theory, it is seriously difficult to determine the initial state in actual experiments. That is because the state is in an extremely-huge configuration space in quantum many-body systems, and we can not obtain the whole information of the state experimentally. It is a fundamental problem in the many-body physics. 

In order to avoid this problem, we utilize the ground state $\big| \text{g} \big>$ of the energy eigen-states, 
\begin{equation}
\left( H + H^{\prime} \right) \big| \text{g} \big> 
= E_{\text{g}}^{\prime} \big| \text{g} \big>, 
\label{Eq-D-2}
\end{equation}
for the initial state, i.e., 
\begin{equation}
\big| \Psi(t_{0}) \big> 
= \big| \text{g} \big>, 
\label{Eq-D-2-2}
\end{equation}
and suddenly remove the additional potential part $H^{\prime}$ in Eq.~(\ref{Eq-D-2}) at $t = t_{0}$. In this scheme, the ground state can approximately be obtained in the experiments of the ultra-cold atomic gases owing to the cooling technique, and we can precisely control the additional potential by using the optical potential. 

In order to study the quantum-mechanical dynamics, we must observe, at least, a time-series of a physical value denoted by $A(t) = \big< \Psi(t) \big| \hat{A} \big| \Psi(t) \big>$ for the corresponding physical value operator $\hat{A}$. The physical value $A(t)$ reflects information of the time-dependent state $\big| \Psi(t) \big>$~\footnote{%
Although the observation can destruct the quantum state, we can obtain the time-series by repeating the experiment from the beginning. That is because the initial state is manually controlled in the present scheme. 
}. 

In the toroidal gases, two of the important physical values must be the toroidal densities $n_{\text{b}}(\theta, t)$ and $n_{\text{f}}(\theta, t)$ defined as 
\begin{equation}
n_{\alpha}(\theta, t) 
= \int_{0}^{\infty} d{r_{c}} r_{c} \int_{- \infty}^{\infty} d{r_{z}} \big< \Psi(t) \big| {\psi_{\alpha}^{(P)}}^{\dagger}(\vec{r}) \psi_{\alpha}^{(P)}(\vec{r}) \big| \Psi(t) \big> 
\label{Eq-D-3}
\end{equation}
for $\alpha = \text{b}$ and $\text{f}$ in the limited configuration space. These distributions directly reflect the diagonal elements of the one-body density matrices in the spatial representation and can experimentally be observed in the optical absorption imaging method.

The toroidal densities $n_{\text{b}}(\theta, t)$ and $n_{\text{f}}(\theta, t)$ in Eq.~(\ref{Eq-D-3}) are reduced to 
\begin{equation}
n_{\text{b}}(\theta, t) 
\approx \sum_{\kappa = - \infty}^{\infty} \sum_{\kappa^{\prime} = - \infty}^{\infty} \frac{e^{i (\kappa^{\prime} - \kappa) \theta}}{2 \pi} \big< \Psi(t) \big| b_{\kappa}^{\dagger} b_{\kappa^{\prime}} \big| \Psi(t) \big> 
\label{Eq-D-4}
\end{equation}
and 
\begin{equation}
n_{\text{f}}(\theta, t) 
\approx \sum_{\kappa = - \infty}^{\infty} \sum_{\kappa^{\prime} = - \infty}^{\infty} \frac{e^{i (\kappa^{\prime} - \kappa) \theta}}{2 \pi} \big< \Psi(t) \big| c_{\kappa}^{\dagger} c_{\kappa^{\prime}} \big| \Psi(t) \big> 
\label{Eq-D-5}
\end{equation}
according to Eqs.~(\ref{Eq-TG-H-11})-(\ref{Eq-TG-H-13}), (\ref{Eq-TG-H-18}), and (\ref{Eq-TG-H-19}). 

Since the toroidal densities $n_{\text{b}}(\theta, t)$ and $n_{\text{f}}(\theta, t)$ are periodic real-valued functions for $\theta$, they can be written as the Fourier series expansion, 
\begin{equation}
n_{\alpha}(\theta, t) 
= \frac{N_{\alpha}}{2 \pi} + \sum_{n = 1}^{\infty} \left[ a_{\alpha n}(t) \cos{(n \theta)} + b_{\alpha n}(t) \sin{(n \theta)} \right] 
\label{Eq-D-6}
\end{equation}
for $\alpha = \text{b}$ and $\text{f}$. Here the constant terms in Eq.~(\ref{Eq-D-6}) proportional to the particle numbers $N_{\text{b}}$ and $N_{\text{f}}$ of the bosons and fermions, respectively, because of the definition of the particle numbers, 
\begin{equation}
N_{\alpha} 
= \int_{0}^{2 \pi} d{\theta}~ n_{\alpha}(\theta, t). 
\label{Eq-D-7}
\end{equation}
Here we assume that the particles mostly exist in the limited configuration space, and dissipation to the residual space is neglected. 

In the present paper, we consider the even functions, $b_{\alpha n}(t) = 0$, and focus on the lowest term, 
\begin{equation}
a_{\alpha 1}(t) 
\equiv \frac{N_{\alpha}}{\pi} d_{\alpha}(t), 
\label{Eq-D-8}
\end{equation}
where we introduce the deformation parameters 
\begin{equation}
d_{\alpha}(t) 
= \frac{1}{N_{\alpha}} \int_{0}^{2 \pi} d{\theta}~ n_{\alpha}(\theta, t) \cos{\theta} 
\label{Eq-D-9}
\end{equation}
for $\alpha = \text{b}$ and $\text{f}$. Here the parity of the toroidal densities is determined by that of the initial state formed from the additional potentials in Eq.~(\ref{Eq-D-2}) because the original Hamiltonian $H$ has the parity symmetry. In addition, the main excitation of the density fluctuation is also determined by the additional potentials because $H$ also has the rotational symmetry. The lowest term must be the easiest mode to excite the toroidal gases. 

In order to excite such a deformation, we prepare the additional potential part $H^{\prime}$ in Eq.~(\ref{Eq-D-2}) as 
\begin{equation}
H^{\prime} 
= \sum_{\alpha = \text{b}, \text{f}} \int d{\vec{r}}~ {\psi_{\alpha}^{(P)}}^{\dagger}(\vec{r}) V_{\alpha} \cos{\theta} \psi_{\alpha}^{(P)}(\vec{r}) 
\label{Eq-D-10}
\end{equation}
with the small potential strengths $V_{\text{b}}$ and $V_{\text{f}}$. It is reduced to 
\begin{equation}
H^{\prime} 
= V_{\text{b}} \hat{d}_{\text{b}} + V_{\text{f}} \hat{d}_{\text{f}}
\label{Eq-D-11}
\end{equation}
with the deformation operators 
\begin{equation}
\hat{d}_{\text{b}} 
\equiv \frac{1}{2} \sum_{\kappa = -\infty}^{\infty} \left( b_{\kappa}^{\dagger} b_{\kappa - 1} + b_{\kappa - 1}^{\dagger} b_{\kappa} \right) 
\label{Eq-D-12}
\end{equation}
and 
\begin{equation}
\hat{d}_{\text{f}} 
\equiv \frac{1}{2} \sum_{\kappa = -\infty}^{\infty} \left( c_{\kappa}^{\dagger} c_{\kappa - 1} + c_{\kappa - 1}^{\dagger} c_{\kappa} \right) 
\label{Eq-D-13}
\end{equation}
according to Eqs.~(\ref{Eq-TG-H-11})-(\ref{Eq-TG-H-13}), (\ref{Eq-TG-H-18}), and (\ref{Eq-TG-H-19}). In fact, the additional potentials in Eq.~(\ref{Eq-D-10}) deform the initial state with the even parity, i.e., $d_{\alpha}(t_{0}) \ne 0$ and $b_{\alpha n}(t) = 0$ in Eq.~(\ref{Eq-D-6}). 

Since 
\begin{equation}
d_{\alpha}(t) 
= \big< \Psi(t) \big| \hat{d}_{\alpha} \big| \Psi(t) \big> 
\label{Eq-D-14}
\end{equation}
for $\alpha = \text{b}$ and $\text{f}$, the energy eigen-value problem for the initial state, $\big| \Psi(t_{0}) \big> = \big| \text{g} \big>$, in Eq.~(\ref{Eq-D-2}) can also be interpreted as an energy minimum problem for $H$, not for $(H + H^{\prime})$, under constrained conditions for $d_{\text{b}}(t_{0})$ and $d_{\text{f}}(t_{0})$ in the Lagrange multiplier method, where $V_{\text{b}}$ and $V_{\text{f}}$ correspond to the Lagrange multipliers for $d_{\text{b}}(t_{0})$ and $d_{\text{f}}(t_{0})$, respectively.  

By using the complete set of the energy eigen-states $\big| n \big>$ for $H$, 
\begin{equation}
H \big| n \big> 
= E_{n} \big| n \big> 
\label{Eq-D-15}
\end{equation}
and 
\begin{equation}
\sum_{n = 0}^{\infty} \big| n \big> \big< n \big| 
= 1, 
\label{Eq-D-16}
\end{equation}
we obtain 
\begin{equation}
d_{\alpha}(t) 
= \sum_{n = 0}^{\infty} \sum_{m = 0}^{\infty} e^{i (E_{n} - E_{m}) t} C_{\alpha n m} 
\label{Eq-D-17}
\end{equation}
with the coefficients 
\begin{equation}
C_{\alpha n m} 
\equiv \big< \text{g} \big| n \big> \big< n \big| \hat{d}_{\alpha} \big| m \big> \big< m \big| \text{g} \big> 
\label{Eq-D-18}
\end{equation}
according to Eqs.~(\ref{Eq-D-1}) and (\ref{Eq-D-2-2}). Thus we can calculate the time-development of $d_{\alpha}(t)$ from the information of the energy eigen-values $E_{n}$ and eigen-states $\big| n \big>$. Note that the deformation parameters $d_{\alpha}(t)$ in Eq.~(\ref{Eq-D-17}) are real numbers, $d_{\alpha}^{*}(t) = d_{\alpha}(t)$, because of the physical value condition, $\hat{d}_{\alpha}^{\dagger} = \hat{d}_{\alpha}$ and $C_{\alpha m n}^{*} = C_{\alpha n m}$. 

In the present paper, we predict the time-series of $d_{\text{b}}(t_{n})$ and $d_{\text{f}}(t_{n})$ from the initial time $t_{0}$ ($\equiv 0$) to a final time $t_{\max}$ as the regular interval data, i.e., 
\begin{equation}
t_{n} 
= n \Delta 
\label{Eq-D-19}
\end{equation}
for the sampling number $n = 0, 1, \dots, (N - 1)$ and interval time $\Delta \equiv t_{\max} / (N - 1)$, and calculate the discrete Fourier transform, 
\begin{equation}
D_{\alpha}(\omega_{m}) 
= \Delta \sum_{n = 0}^{N - 1} e^{- i \omega_{m} t_{n}} d_{\alpha}(t_{n}) 
\label{Eq-D-20}
\end{equation}
for $\alpha = \text{b}$ and $\text{f}$, $\omega_{m} \equiv 2 \pi m / (N \Delta)$, and $m = (- N / 2), (- N / 2 + 1), \dots, (N / 2)$. 

Note that the Fourier transform 
\begin{equation}
D_{\alpha}(\omega) 
\equiv \int_{- \infty}^{\infty} d{t}~ e^{- i \omega t} d_{\alpha}(t) 
\label{Eq-D-21}
\end{equation}
is reduced to 
\begin{equation}
D_{\alpha}(\omega) 
= \sum_{n = 0}^{\infty} \sum_{m = 0}^{\infty} 2 \pi C_{\alpha n m} \delta{(E_{n} - E_{m} - \omega)} 
\label{Eq-D-22}
\end{equation}
according to Eq.~(\ref{Eq-D-17}). In the contiguous limit, i.e., $t_{\max} \to \infty$ and $\Delta \to 0$, the discrete Fourier transform in Eq.~(\ref{Eq-D-20}) gets close to the Fourier transform in Eq.~(\ref{Eq-D-22}). 

In general, how to visualize a large number of $\delta$ functions, e.g., those in Eq.~(\ref{Eq-D-22}), is a technical problem in quantum many-body physics. In fact, the selection of the visualization methods obviously affects detailed profiles of the graphics of the calculational results. 

In the present paper, we select the method of the discrete Fourier transform in Eq.~(\ref{Eq-D-20}) because it can reproduce that in actual experiments, where the observation data can be obtained as the time-series of $d_{\text{b}}(t_{n})$ and $d_{\text{f}}(t_{n})$.

\subsection{Calculational schemes \label{SubSec-calculation}}
Here we explain the calculational schemes used in the present paper. First we explain the numerical exact diagonalization method as the main scheme in the present paper. Second we also explain the HF approximation used for comparison.

\subsubsection{Exact diagonalization method \label{SubSec-calculation-exact}}

In the exact diagonalization method, we numerically solve the energy eigen-value problems in Eqs.~(\ref{Eq-D-2}) and (\ref{Eq-D-15}) by diagonalizing the Hamiltonian matrix in a limited space in the matrix representation, where the diagonalization is exactly computed except for numeric errors. The quantum-mechanical dynamics can be calculated from the eigen-values and eigen-states according to Eqs.~(\ref{Eq-D-17}) and (\ref{Eq-D-18}).

As an advantage of this method, we can obtain exact solutions except for numeric errors and limitation of the matrix space. On the other hand, as a disadvantage of this method, we can not demonstrate the calculation for large-number and strongly-interacting systems, where the needed matrix space becomes extremely huge and overflows from our computer memories. 

In order to demonstrate the calculation, we decide the matrix space as follows. 

First we make the base vectors by using the Fock space for the boson and fermion operators $b_{\kappa}$ and $c_{\kappa}$, where the particle numbers $N_{\text{b}}$ and $N_{\text{f}}$ are fixed. These base vectors work well except for the highly-deformed and/or strongly-interacting gases as a result. 

Second we limit the range of $\kappa$ as $\left| \kappa \right| \le \kappa_{\text{b} \max}$ for the bosons and $\left| \kappa \right| \le \kappa_{\text{f} \max}$ for the fermions. This limit can especially affect the interaction energies for the contact-type interactions in Eqs.~(\ref{Eq-TG-H-28}) and (\ref{Eq-TG-H-29}); However, these effects on the dynamics are visually small except for the case of the strongly-interacting gases. 

Third we also limit the matrix space by eliminating the base vectors with the higher kinetic excitation energies than a cut-off energy $e_{\text{c}}$. Here the kinetic excitation energy of one of the base vectors means difference of the expectation values of $(H_{\text{b}} + H_{\text{f}})$ in Eqs.~(\ref{Eq-TG-H-22}) and (\ref{Eq-TG-H-23}) between the base vector and ground state of the non-interacting gases. This limit is valid when the cut-off energy is much greater than the typical scales of the excitation and correlation energies, where the correlation energy is defined as difference between the interaction energy and mean-field energy. 
\subsubsection{The Hartree-Fock approximation \label{SubSec-calculation-HF}}
The HF mean-field approximation must be one of the most popular schemes to treat the ultra-cold atomic gases. Here, as a naming convention, the ``Hartree-Fock'' approximation is used for fermions, and the ``Hartree'' approximation is used for bosons. In the present paper, we use the ``Hartree-Fock'' approximation for the boson-fermion mixtures as the unified name. 

In the HF approximation, the configuration space of the state $\big| \Psi(t) \big>$ is limited to that of the HF state denoted by 
\begin{equation}
\big| \Psi_{\text{HF}}(t) \big> 
= \frac{\left[ \tilde{b}^{\dagger}(t) \right]^{N_{\text{b}}}}{\sqrt{N_{\text{b}}!}} \prod_{n = 1}^{N_{\text{f}}} \tilde{c}_{n}^{\dagger}(t) \big| 0 \big> 
\label{Eq-calculation-HF-1}
\end{equation}
with the vacuum $\big| 0 \big>$ and redefined boson and fermion operators 
\begin{equation}
\tilde{b}(t) 
\equiv \sum_{\kappa = - \infty}^{\infty} \varphi_{\kappa}^{*}(t) b_{\kappa} 
\label{Eq-calculation-HF-2}
\end{equation}
and 
\begin{equation}
\tilde{c}_{n}(t) 
\equiv \sum_{\kappa = - \infty}^{\infty} u_{n \kappa}^{*}(t) c_{\kappa}, 
\label{Eq-calculation-HF-3}
\end{equation}
where the single-particle wave-functions $\varphi_{\kappa}(t)$ and $u_{n \kappa}(t)$ obey 
\begin{equation}
\sum_{\kappa = - \infty}^{\infty} \varphi_{\kappa}^{*}(t) \varphi_{\kappa}(t) 
= 1, 
\label{Eq-calculation-HF-4}
\end{equation}
\begin{equation}
\sum_{\kappa = - \infty}^{\infty} u_{n \kappa}^{*}(t) u_{n^{\prime} \kappa}(t) 
= \delta_{n, n^{\prime}}, 
\label{Eq-calculation-HF-5}
\end{equation}
and 
\begin{equation}
\sum_{n = 1}^{\infty} u_{n \kappa}^{*}(t) u_{n \kappa^{\prime}}(t) 
= \delta_{\kappa, \kappa^{\prime}}. 
\label{Eq-calculation-HF-6}
\end{equation}

In the HF approximation, information of the many-body state is abbreviated into the single-particle wave-functions $\varphi_{\kappa}(t)$ and $u_{n \kappa}(t)$ in Eqs.~(\ref{Eq-calculation-HF-2}) and (\ref{Eq-calculation-HF-3}) related to the one-body density matrices. As a result, the many-body correlations beyond the one-body density matrices are neglected in this scheme. The validity of this scheme is discussed in the following sections. 

The single-particle wave-functions $\varphi_{\kappa}(t)$ and $u_{n \kappa}(t)$ in Eqs.~(\ref{Eq-calculation-HF-2}) and (\ref{Eq-calculation-HF-3}) are determined by the variational minimization of the action 
\begin{equation}
S_{\text{HF}} 
\equiv \int d{t}~ \big< \Psi_{\text{HF}}(t) \big| i \frac{d}{d{t}} - H \big| \Psi_{\text{HF}}(t) \big> 
\label{Eq-calculation-HF-7}
\end{equation}
for the dynamics in Eq.~(\ref{Eq-D-1}) and of the energy 
\begin{equation}
E_{\text{HF}}^{\prime} 
\equiv \big< \Psi_{\text{HF}}(t_{0}) \big| H + H^{\prime} \big| \Psi_{\text{HF}}(t_{0}) \big> 
\label{Eq-calculation-HF-8}
\end{equation}
for the ground and initial state in Eq.~(\ref{Eq-D-2}) with the normalization constraint in Eqs.~(\ref{Eq-calculation-HF-4}) and (\ref{Eq-calculation-HF-5}). 

After all, for the dynamics in Eq.~(\ref{Eq-D-1}), we obtain the coupled time-dependent HF equations, 
\begin{eqnarray}
&& 
i \frac{d{\varphi_{\kappa}}}{d{t}}(t) 
= \sum_{\kappa^{\prime} = - \infty}^{\infty} \bigg[ \frac{\kappa^{2} \delta_{\kappa, \kappa^{\prime}}}{2 m_{\text{b}} R^{2}} + \frac{g_{\text{bb}}}{2 \pi R} \frac{N_{\text{b}} - 1}{N_{\text{b}}} \rho_{\kappa^{\prime} - \kappa}^{(\text{b})(\text{HF})}(t) 
\nonumber \\ 
&& 
+ \frac{g_{\text{bf}}}{2 \pi R} \rho_{\kappa^{\prime} - \kappa}^{(\text{f})(\text{HF})}(t) \bigg] \varphi_{\kappa^{\prime}}(t) 
\label{Eq-calculation-HF-9}
\end{eqnarray}
and 
\begin{equation}
i \frac{d{u_{n \kappa}}}{d{t}}(t) 
= \sum_{\kappa^{\prime} = - \infty}^{\infty} \bigg[ \frac{\kappa^{2} \delta_{\kappa, \kappa^{\prime}}}{2 m_{\text{f}} R^{2}} 
+ \frac{g_{\text{bf}}}{2 \pi R} \rho_{\kappa^{\prime} - \kappa}^{(\text{b})(\text{HF})}(t) \bigg] u_{n \kappa^{\prime}}(t) 
\label{Eq-calculation-HF-10}
\end{equation}
with the momentum distributions 
\begin{equation}
\rho_{\kappa}^{(\text{b})(\text{HF})}(t) 
\equiv N_{\text{b}} \sum_{\kappa^{\prime} = - \infty}^{\infty} \varphi_{\kappa^{\prime}}^{*}(t) \varphi_{\kappa^{\prime} - \kappa}(t) 
\label{Eq-calculation-HF-11}
\end{equation}
and 
\begin{equation}
\rho_{\kappa}^{(\text{f})(\text{HF})}(t) 
\equiv \sum_{n = 1}^{N_{\text{f}}} \sum_{\kappa^{\prime} = - \infty}^{\infty} u_{n \kappa^{\prime}}^{*}(t) u_{n (\kappa^{\prime} - \kappa)}(t). 
\label{Eq-calculation-HF-12}
\end{equation}
Note that, if we replace the coefficient $(N_{\text{b}} - 1) / N_{\text{b}}$ in Eq.~(\ref{Eq-calculation-HF-9}) with unity by neglecting $N_{\text{b}}^{- 1}$, Eq.~(\ref{Eq-calculation-HF-9}) corresponds to the Gross-Pitaevskii equation. 

For the ground and initial state in Eq.~(\ref{Eq-D-2}) at $t = t_{0}$, we obtain the coupled HF equations for $\varphi_{\kappa}(t_{0})$ and $u_{n \kappa}(t_{0})$, 
\begin{eqnarray}
&& 
\varepsilon^{(\text{b})} \varphi_{\kappa} 
= \sum_{\kappa^{\prime} = - \infty}^{\infty} \bigg[ \frac{\kappa^{2} \delta_{\kappa, \kappa^{\prime}}}{2 m_{\text{b}} R^{2}} + V_{\text{b}} \frac{\delta_{(\kappa - 1), \kappa^{\prime}} + \delta_{\kappa, (\kappa^{\prime} - 1)}}{2} 
\nonumber \\ 
&& 
+ \frac{g_{\text{bb}}}{2 \pi R} \frac{N_{\text{b}} - 1}{N_{\text{b}}} \rho_{\kappa^{\prime} - \kappa}^{(\text{b})(\text{HF})}(t_{0}) + \frac{g_{\text{bf}}}{2 \pi R} \rho_{\kappa^{\prime} - \kappa}^{(\text{f})(\text{HF})}(t_{0}) \bigg] \varphi_{\kappa^{\prime}} 
\label{Eq-calculation-HF-13}
\end{eqnarray}
and 
\begin{eqnarray}
&& 
\varepsilon_{n}^{(\text{f})} u_{n \kappa} 
= \sum_{\kappa^{\prime} = - \infty}^{\infty} \bigg[ \frac{\kappa^{2} \delta_{\kappa, \kappa^{\prime}}}{2 m_{\text{f}} R^{2}} + V_{\text{f}} \frac{\delta_{(\kappa - 1), \kappa^{\prime}} + \delta_{\kappa, (\kappa^{\prime} - 1)}}{2} 
\nonumber \\ 
&& 
+ \frac{g_{\text{bf}}}{2 \pi R} \rho_{\kappa^{\prime} - \kappa}^{(\text{b})(\text{HF})}(t_{0}) \bigg] u_{n \kappa^{\prime}} 
\label{Eq-calculation-HF-14}
\end{eqnarray}
with the single-particle energies $\varepsilon^{(\text{b})}$ and $\varepsilon_{n}^{(\text{f})}$. In the HF ground state, according to Eq.~(\ref{Eq-calculation-HF-1}), all of the bosons occupy the lowest single-particle state, and the fermions exhibit the Fermi degeneracy. 

In the small amplitude excitations from the HF ground state, the HF approximation agree with the random phase approximation (RPA)~\cite{RS, Thouless} described in appendix~\ref{APP_2}. In the present framework, the amplitudes of the excitations are not limited in the small amplitude excitations, and we thus use the HF approximation. 

In addition, instability of the HF ground state can be verified by RPA because the instability can be seen in the small amplitude excitations as explained in appendix~\ref{APP_2}. The instability condition for the HF ground state is discussed in the next section. 
%
%
\section{Static properties \label{Sec-static}}
%
%
We here roughly explain static properties of the toroidal gases in the ground state as an important background knowledge to study the dynamics. In Sec.~\ref{Sec-static-A}, we explain effects of the quantum statistics and inter-particle interactions on the toroidal gases. In Sec.~\ref{Sec-static-B}, we also explain the static properties of the non-deformed toroidal gases without the deformation potentials and consider the instability condition of the HF ground state. 
\subsection{Effects of quantum statistics and interactions \label{Sec-static-A}}
In the non-interacting gases, all of the bosons occupy the lowest single-particle state, and the fermions exhibit the Fermi degeneracy, in common with the HF ground state. In fact, the ground state of the non-interacting gases agrees with the HF ground state. 

If we add the same deformation potential for the bosons and fermions to the non-interacting gases as described in Eq.~(\ref{Eq-D-2}), the bosons exhibit larger deformation than that of the fermions. That is because the fermions need more energy to deform the gas owing to the Fermi pressure. It is a typical difference between the bosons and fermions. 

Similarly the fermions are more insensitive to the inter-particle interactions than the bosons owing to the Pauli exclusion principle in two mechanisms as follows. First the fermion-fermion interaction is neglected in the model in Eq.~(\ref{Eq-TG-H-32}) because of the Pauli blocking effect. Second the fermions has more kinetic energy than the bosons because of the Fermi degeneracy, and the motion of the fermions relatively reduces effects of the interactions. As a result, in a rough picture, the interactions chiefly affect the bosons, and the fermions reluctantly obey the bosons owing to the boson-fermion interaction. 

Although the three-dimensional gases can exhibit the collapse phenomena owing to the inter-atomic attractions, the quasi-one-dimensional gases can not collapse in theory. That is because the quantum-mechanical kinetic energy (proportional to $L^{-2}$ with the typical length size $L$ of the gases for the small $L$) overcomes the attraction energy (proportional to $L^{-1}$) in the quasi-one-dimensional gases. Of course, it is a theoretical artifact derived from the limited configuration space for the quasi-one-dimensional model, and the gases can collapse in the full configuration space because of the three-dimensional attraction energy (proportional to $L^{-3}$). The theoretical stability of the quasi-one-dimensional gases, however, indicates realization of long-lived meta-stable states with the attractions in actual experiments. In fact, such states like the bright soliton are realized in the quasi-one-dimensional bosons with the attractions in the ultra-cold atomic gases.

In the present paper, we thus theoretically consider not only the inter-atomic repulsions but also the attractions. 
\subsection{The non-deformed systems \label{Sec-static-B}}
In the non-deformed toroidal gases without the deformation potentials, the ground state must have the rotational symmetry in the finite quantum systems, where spontaneous symmetry breaking is forbidden. In fact, the ground state calculated in the exact diagonalization method has the rotational symmetry. 

Since the HF ground state is merely an approximation for the exact ground state, it may be unstable for the density fluctuation, and the instability of the HF ground state is usually related to symmetry breaking in general. Here it should be noted that the symmetry breaking for the mean-field theory is strictly different from the spontaneous symmetry breaking in quantum field theory for the infinite matter systems. 

The instability condition of the non-deformed HF ground state is obtained from the lowest excitation energy of the small amplitude excitations from the HF ground state, where the time-dependent HF approximation is reduced to RPA as described in appendix~\ref{APP_2}. In fact, the excitation energy can go into an imaginary number, where the corresponding fluctuation grows in time and indicates the instability. 

The instability appears only in the non-deformed systems and disappears in the deformed systems with the deformation potentials. That is because the instability is linked to the symmetry breaking and the quasi-one-dimensional gases can not collapse in theory. 

The instability condition for the non-deformed systems can also be obtained from behaviors of the deformation parameters in the HF ground state in the deformed systems with the small additional deformation potentials. Under the instability condition, the deformation parameters suddenly grow by addition of the infinitesimal potentials, and the gases exhibit the soliton-like profiles~\cite{BF_soliton}, where the gases are spatially localized in order to balance the interaction and kinetic energies. 

%
%
\begin{figure}[ht]
\begin{center}
  \begin{tabular}{c}
       \includegraphics[scale=0.4,angle=-90]{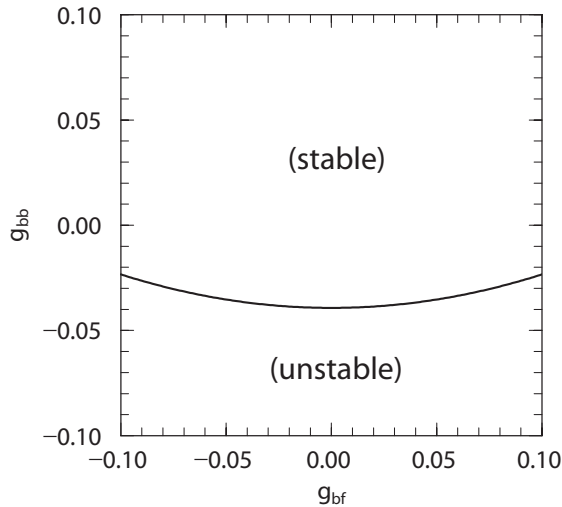}
  \end{tabular}
\end{center}
\caption{The instability condition of the non-deformed HF ground state. The setting parameters and scales are same as those in Sec.~\ref{Sec-dynamic} except for $V_{\text{b}} = V_{\text{f}} = 0$.} 
\label{Fig-pd-rpa}
\end{figure}
%
%

In Fig.~\ref{Fig-pd-rpa}, we show the instability condition as a typical demonstration, where the setting parameters and scales are same as those in Sec.~\ref{Sec-dynamic} except for $V_{\text{b}} = V_{\text{f}} = 0$. 

As shown in Fig.~\ref{Fig-pd-rpa}, the boson-boson attraction can make the instability and induce the bright soliton for the bosons as a main mechanism of the instability. In particular, when $g_{\text{bf}} =0$, the instability occurs at 
\begin{equation}
g_{\text{bb}} 
\le - \frac{\pi}{2 m_{\text{b}} R (N_{\text{b}} - 1)} 
\equiv g_{\text{bb}}^{(\text{c})}. 
\label{Sec-static-B-1}
\end{equation}
This condition agrees with that in the time-dependent Gross-Pitaevskii equation, i.e., the Bogoliubov analysis, except for the difference term $N_{\text{b}}^{- 1}$ from the HF approximation. 

In Fig.~\ref{Fig-pd-rpa}, the boson-fermion interaction also affects the instability and can induce the phase-separation-like profiles for the repulsion, $g_{\text{bf}} > 0$, and the mixed soliton-like profiles for the attraction, $g_{\text{bf}} < 0$, as the other mechanism of the instability. As an aspect of this mechanism, the instability condition appears symmetrically for the boson-fermion repulsion and attraction as shown in Fig.~\ref{Fig-pd-rpa}. 

Although the instability of the HF ground state is a theoretical artifact in the HF approximation, it is deeply linked to the dynamics beyond the HF approximation as demonstrated and discussed in the following sections. 
\section{Dynamical properties \label{Sec-dynamic}}
In this section, we demonstrate the dynamics in some typical situations. In Sec.~\ref{SubSec-parameter}, we first explain the setting parameters and scales needed to demonstrate the dynamics. In Sec.~\ref{SubSec-dynamics}, we show the calculational results. 
\subsection{Setting parameters and scales \label{SubSec-parameter}}
Since one of the purpose of the present work is to study effects of the quantum statistics on the dynamics, we select the symmetric parameters for the bosons and fermions, i.e., $m_{\text{b}} = m_{\text{f}}$, $V_{\text{b}} = V_{\text{f}}$, and $N_{\text{b}} = N_{\text{f}}$. In this case, both of the particles exhibit the same behaviors except for the effects of the quantum statistics. 

We also determine the scales as follows without loss of generality. In addition to $\hbar = 1$ set in the above sections, we also select $m_{\text{b}} = m_{\text{f}} = 1$ for the mass scale and $R / (N_{\text{b}} + N_{\text{f}}) = 1$ for the length scale. Note that we here select the inverse of the total density for the length scale, instead of $R$, because the total density is kept in the thermodynamical limit, differently from $R$. In these scales, the typical absolute values of the coupling constants, $\left| g_{\text{bb}} \right|$ and $\left| g_{\text{bf}} \right|$, in actual experiments become $10^{- (1 \sim 3)}$ depending on detailed setup of experimental situations. 

We concretely set $N_{\text{b}} = N_{\text{f}} = 5$ and $V_{\text{b}} = V_{\text{f}} = - 0.001$ in the following demonstration. Here the precise value of $V_{\text{b}}$ and $V_{\text{f}}$ hardly affect the calculational results. The numbers $N_{\text{b}}$ and $N_{\text{f}}$ are chiefly determined by the limitation of our computer memories. In addition, it is theoretically favorable for us to select an odd number for $N_{\text{f}}$. That is because the Fermi levels, which are degenerate double except for the ground state owing to the parity symmetry, are fully occupied for the odd number; Otherwise the single fermion on the Fermi levels exhibits a precise beat structure in the density distribution, and, after all, the calculational results for the odd number are clearer than those for the even number. 

In practice, the other parameters needed to demonstrate the dynamics are determined as follows. The cut-off momenta are determined as $\kappa_{\text{b} \max} = 4$ for the bosons and $\kappa_{\text{f} \max} = 5$ for the fermions, where the imbalance for the bosons and fermions is due to the Fermi degeneracy only for the fermions. The cut-off energy is determined as $e_{\text{c}} = 0.14$. Here the cut-off energy is much greater than the typical scales of the effective excitation energies, which are shown in the results in the following figures, and absolute values of the correlation energies, which are comparable with or smaller than that of the mean-field energy, $\big| e_{\text{m.f.}} \big| \sim \big| g_{\text{bb}}^{(\text{c})} \big| N_{\text{b}}^{2} / (4 \pi R) \approx 0.0078$, in the weakly-interacting gases studied below. 

In addition, we also determine the sampling number $n$ and maximum time $t_{\max}$ in order to obtain the discrete Fourier transforms $D_{\text{b}}$ and $D_{\text{f}}$ in Eq.~(\ref{Eq-D-20}) as $N = 2^{11}$ and $t_{\max} = 20 \times N$. Here it should be noted that detailed profiles of the discrete Fourier transforms depend on the sampling parameters, $N$ and $t_{\max}$, and these parameters may have some practical limitations in experiments, e.g., the life-time of the gases, thermal damping, and so on. Thus the detailed profiles may not be serious in the present paper, and we here focus on the qualitative profiles robust over a large region of the sampling parameters.

\subsection{Calculational results \label{SubSec-dynamics}}
Here we show the calculational results when $g_{\text{bf}} \leq 0$ in Sec.~\ref{SubSec-att} and when $g_{\text{bf}} > 0$ in Sec.~\ref{SubSec-rep}. Some theoretical interpretations of the results are discussed in the next section.  

In the present paper, we especially focus on the boson-boson attraction case, $g_{\text{bb}} < 0$. That is because the boson-boson attraction induces a qualitative change of the dynamics around $g_{\text{bb}} = g_{\text{bb}}^{(\text{c})}$ related to the instability of the HF ground state in Fig.~\ref{Fig-pd-rpa} as demonstrated below.

\subsubsection{The boson-fermion attraction case \label{SubSec-att}}
\begin{figure}[ht]
\begin{center}
\includegraphics[width=8.0cm]{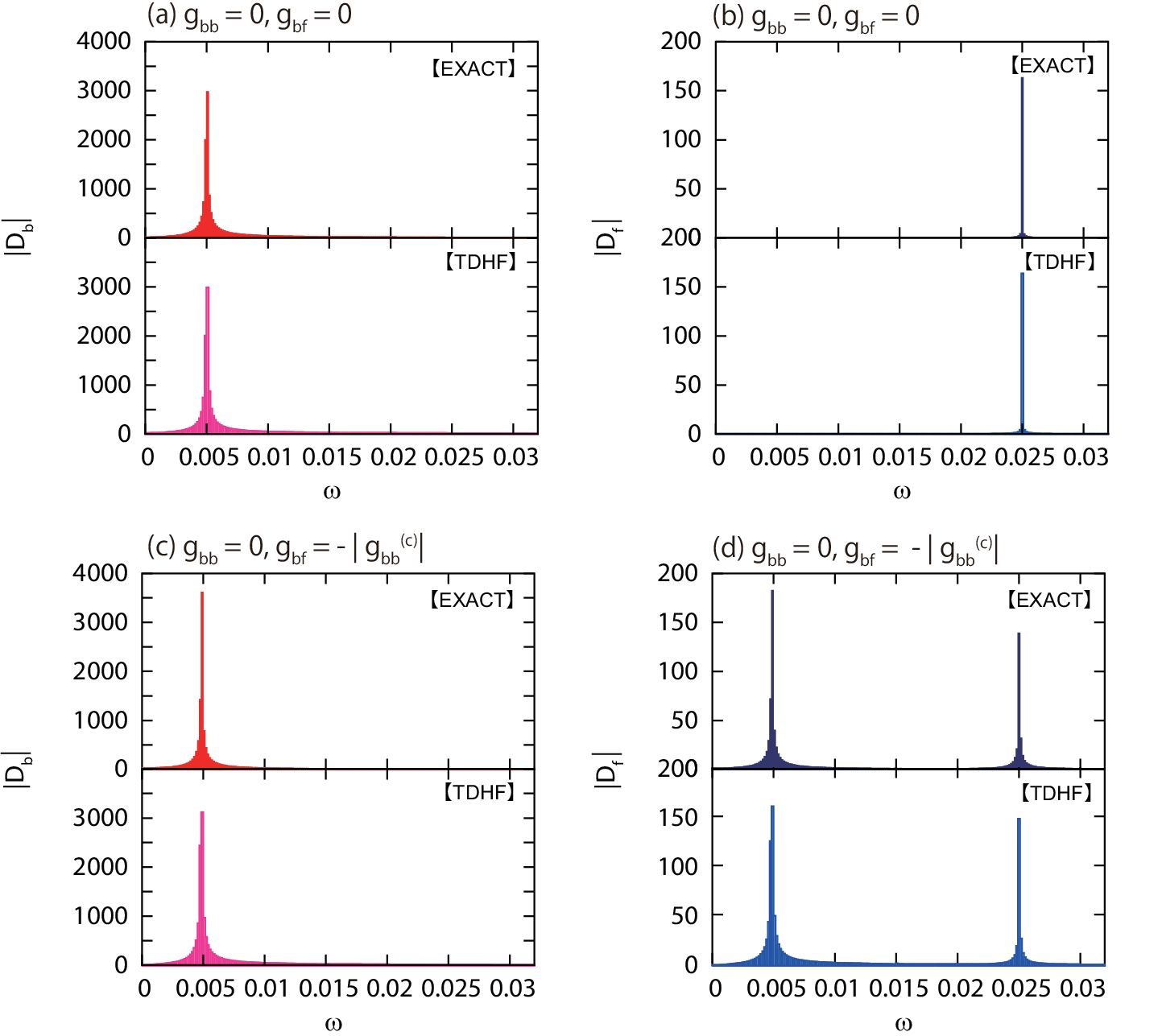}
\end{center}
\caption{(Color online) The absolute values of the discrete Fourier transforms,  $\left| D_{\text{b}} \right|$ [(a) (c)] and $\left| D_{\text{f}} \right|$ [(b) (d)], when $g_{\text{bb}} = 0$ and $g_{\text{bf}} = 0$ [(a) (b)] and $- \big| g_{\text{bb}}^{(\text{c})} \big|$ [(c) (d)]. The top and bottom columns represent the results in the exact diagonalization method and HF approximation, respectively.}
\label{F2}
\end{figure}

\begin{figure}[ht]
\begin{center}
\includegraphics[width=8.0cm]{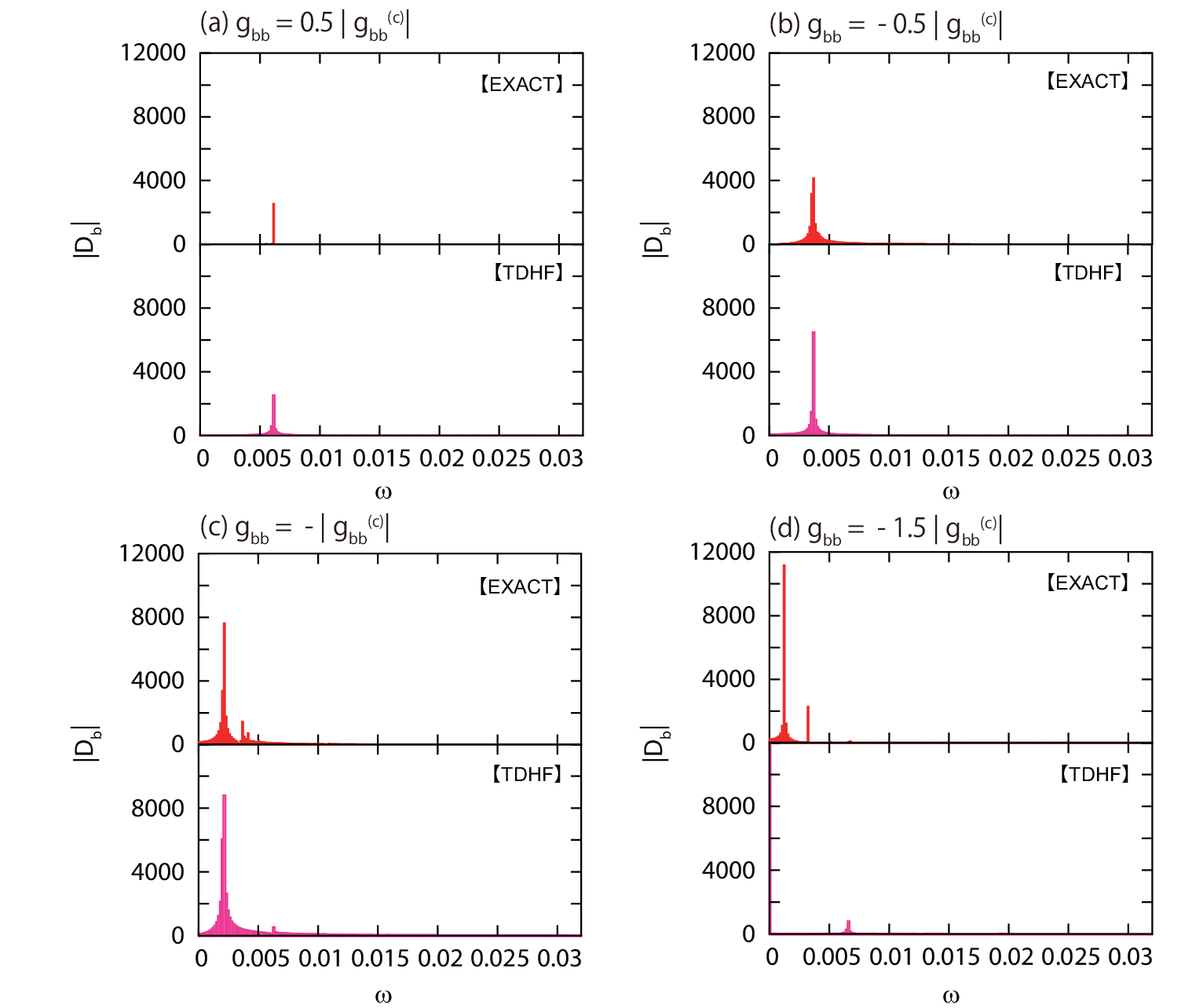}
\end{center}
\caption{(Color online) The absolute values of the discrete Fourier transforms, $\left| D_{\text{b}} \right|$, when $g_{\text{bf}} = 0$ and $g_{\text{bb}} = (1 / 2) \big| g_{\text{bb}}^{(\text{c})} \big|$ (a), $(- 1 / 2) \big| g_{\text{bb}}^{(\text{c})} \big|$ (b), $- \big| g_{\text{bb}}^{(\text{c})} \big|$ (c), and $(- 3 / 2) \big| g_{\text{bb}}^{(\text{c})} \big|$ (d). The top and bottom columns represent the results in the exact diagonalization method and HF approximation, respectively.}
\label{F3}
\end{figure}

\begin{figure}[ht]
\begin{center}
\includegraphics[width=8.0cm]{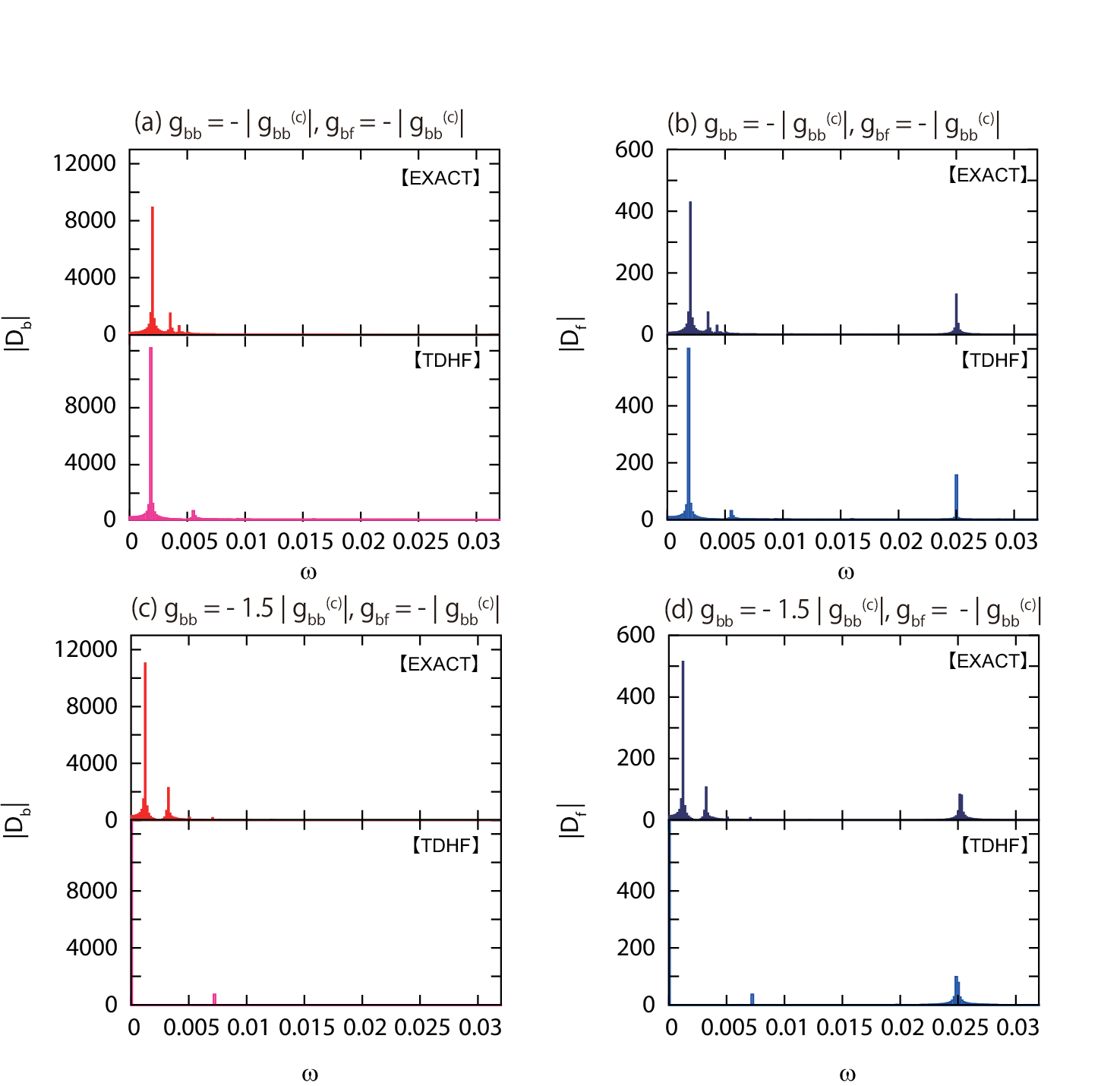}
\end{center}
\caption{(Color online) The absolute values of the discrete Fourier transforms, $\left| D_{\text{b}} \right|$ [(a) (c)] and $\left| D_{\text{f}} \right|$ [(b) (d)], when $g_{\text{bf}} = - \big| g_{\text{bb}}^{(\text{c})} \big|$ and $g_{\text{bb}} = - \big| g_{\text{bb}}^{(\text{c})} \big|$ [(a) (b)] and $(- 3 / 2)\big| g_{\text{bb}}^{(\text{c})} \big|$ [(c) (d)]. The top and bottom columns represent the results in the exact diagonalization method and HF approximation, respectively.}
\label{F4}
\end{figure}

First we show the results for the $g_{\text{bb}} = 0$ case in Fig.~\ref{F2} in order to demonstrate effects of the boson-fermion attraction on the dynamics. Second we show the results for the $g_{\text{bf}} = 0$ case in Fig.~\ref{F3} in order to demonstrate effects of the boson-boson interaction. Third we show the results for the $g_{\text{bb}} < 0$ and $g_{\text{bf}} < 0$ case in Fig.~\ref{F4}. 

%
%

In Fig.~\ref{F2}, we show the absolute values of the discrete Fourier transforms, $\left| D_{\text{b}} \right|$ and $\left| D_{\text{f}} \right|$ in Eq.~(\ref{Eq-D-20}), when $g_{\text{bb}} = 0$. Note that the Fourier transforms are even functions, $D_{\text{b}}(- \omega_{m}) = D_{\text{b}}(\omega_{m})$ and $D_{\text{f}}(- \omega_{m}) = D_{\text{f}}(\omega_{m})$, because of the real-valued $d_{\text{b}}(t)$ and $d_{\text{f}}(t)$, and visually zero in the upper outside of the plotted range in Fig.~\ref{F2} and also in Figs.~\ref{F3}-\ref{F5}. 

In Figs.~\ref{F2}(a) and \ref{F2}(b), the results in $g_{\text{bb}} = g_{\text{bf}} = 0$ exhibit single sharp peaks at $\omega \approx 0.0050$ for the bosons and $\omega \approx 0.0250$ for the fermions, respectively. Here the results in the exact diagonalization method and HF approximation plotted in the top and bottom columns, respectively, are visually same except for the negligible numeric errors. The positions of the peaks agree with the first excitation energies linked to the quantum statistics, i.e., $1 / (2 m_{\text{b}} R^{2}) = 0.005$ for the bosons and $(3^{2} - 2^{2}) / (2 m_{\text{f}} R^{2}) = 0.025$ for the fermions. The difference of the vertical scales for the bosons and fermions indicates that of the oscillation amplitudes determined by the deformation parameters in the initial state. i.e., the effect of the Fermi pressure explained in Sec.~\ref{Sec-static-A}. 

In Figs.~\ref{F2}(c) and \ref{F2}(d), the results in $g_{\text{bf}} = - \big| g_{\text{bb}}^{(\text{c})} \big|$ exhibit almost the same structure in Figs.~\ref{F2}(a) and \ref{F2}(b) except for a new peak only for the fermions. The new peak is exactly analogous to the peak for the bosons in the shape and position. According to the HF approximation in Eqs.~(\ref{Eq-calculation-HF-9}) and (\ref{Eq-calculation-HF-10}), the new peak for the fermions can be induced by the bosons as a forced oscillation owing to the boson-fermion attraction. On the other hand, the effect of the fermions on the bosons is visually small. 

%
%

In Fig.~\ref{F3}, we show $\left| D_{\text{b}} \right|$ when $g_{\text{bf}} = 0$, where $\left| D_{\text{f}} \right|$ is same as that in Fig.~\ref{F2}b. 

In Figs.~\ref{F3}(a) and \ref{F3}(b), the results in $g_{\text{bb}} = (1 / 2) \big| g_{\text{bb}}^{(\text{c})} \big|$ and $(- 1 / 2) \big| g_{\text{bb}}^{(\text{c})} \big|$ exhibit single peaks at $\omega \approx 0.0061$ and $0.0036$, respectively. Here the results in the HF approximation agrees with those in the exact diagonalization method because of the weak interactions. The heights of the peaks in Figs.~\ref{F3}(a), \ref{F2}(a), and \ref{F3}(b) reflect those of the oscillation amplitudes according to the Parseval's theorem. 

In Fig.~\ref{F3}(c), the results at the instability boundary of the HF ground state, $g_{\text{bb}} = g_{\text{bb}}^{(\text{c})}$, exhibit a major peak at $\omega \approx 0.0021$ and minor peaks near $\omega \approx 0.0037$. The HF approximation does not reproduce the minor peaks. 

In Fig.~\ref{F3}(d), the results in $g_{\text{bb}} = (- 3 / 2) \big| g_{\text{bb}}^{(\text{c})} \big|$ exhibit a major peak at $\omega \approx 0.0012$ and a minor peak at $\omega \approx 0.0032$. Here the results in the HF approximation exhibit quite different behaviors from the exact results and indicate the dynamics of the soliton. In fact, the results in the HF approximation exhibit a dominant peak at $\omega = 0$, corresponding to a stationary soliton, and a minor peak at $\omega \approx 0.0066$, corresponding to surface oscillations of the soliton, where the minor peak is also slightly seen in Fig.~\ref{F3}(c) at $\omega \approx 0.0063$.

%
%

In Fig.~\ref{F4}, we show $\left| D_{\text{b}} \right|$ and $\left| D_{\text{f}} \right|$ when $g_{\text{bf}} = - \big| g_{\text{bb}}^{(\text{c})} \big|$ and $g_{\text{bb}} < 0$. These situations correspond to those in Figs.~\ref{F2}(b), \ref{F3}(c), and \ref{F3}(d) except for the boson-fermion attraction. 

Here both of the results in $g_{\text{bb}} = - \big| g_{\text{bb}}^{(\text{c})} \big|$ and $(- 3 / 2) \big| g_{\text{bb}}^{(\text{c})} \big|$ exhibit almost the same structure in Figs.~\ref{F2}(b), \ref{F3}(c), and \ref{F3}(d) except for new peaks only for the fermions in analogy with the case in Fig.~\ref{F2}. The mechanism of the forced oscillation must not be affected by the boson-boson interaction as demonstrated here.

\subsubsection{The boson-fermion repulsion case \label{SubSec-rep}}

Here we show the results when $g_{\text{bf}} > 0$ and comment on symmetric properties for the boson-fermion attraction and repulsion. 

%
\begin{figure}[ht]
\begin{center}
\includegraphics[width=8.0cm]{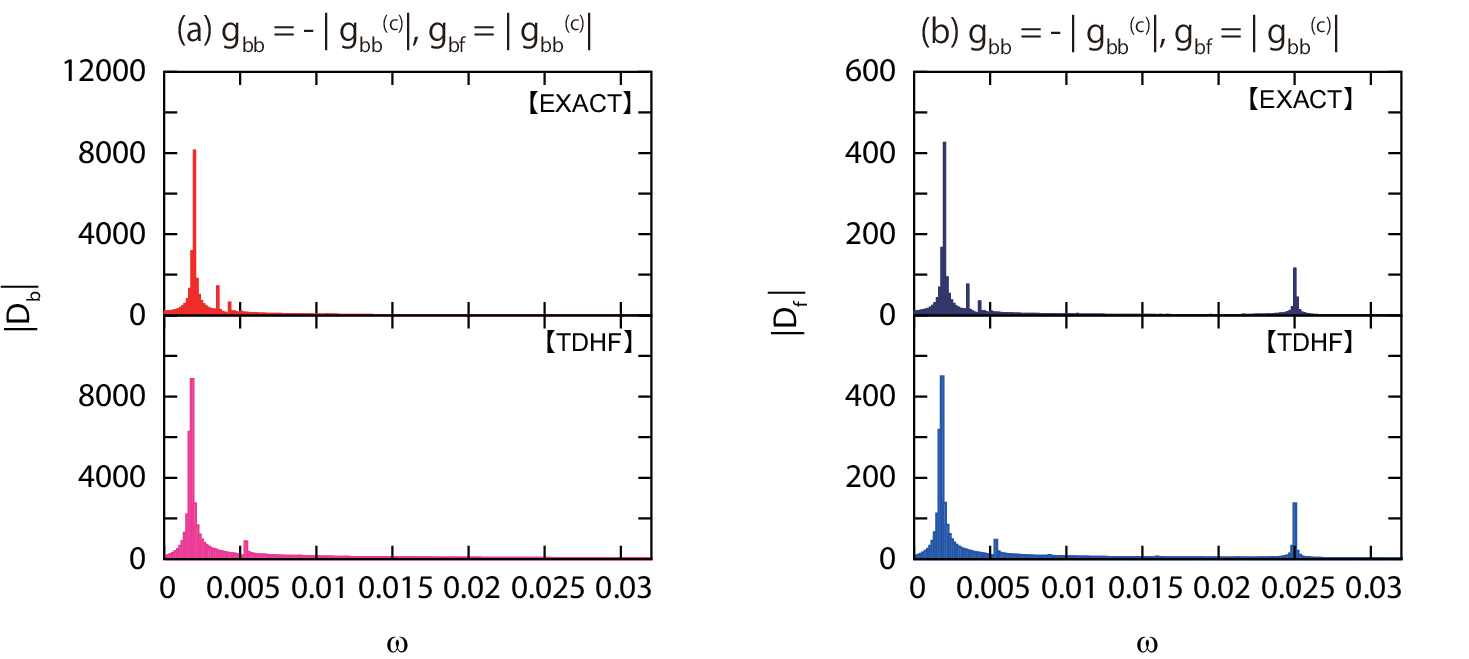}
\end{center}
\caption{(Color online) The absolute values of the discrete Fourier transforms, $\left| D_{\text{b}} \right|$ (a) and $\left| D_{\text{f}} \right|$ (b), when $g_{\text{bf}} = \big| g_{\text{bb}}^{(\text{c})} \big|$ and $g_{\text{bb}} = - \big| g_{\text{bb}}^{(\text{c})} \big|$. The top and bottom columns represent the results in the exact diagonalization method and HF approximation, respectively.}
\label{F5}
\end{figure}
%

In Fig.~\ref{F5}, we show $\left| D_{\text{b}} \right|$ and $\left| D_{\text{f}} \right|$ when $g_{\text{bf}} = \big| g_{\text{bb}}^{(\text{c})} \big|$ and $g_{\text{bb}} = - \big| g_{\text{bb}}^{(\text{c})} \big|$. This situation agrees with that in Figs.~\ref{F4}(a) and \ref{F4}(b) except for the sign of $g_{\text{bf}}$. 

As shown in Figs.~\ref{F5}, \ref{F4}(a), and \ref{F4}(b), $\left| D_{\text{b}} \right|$ and $\left| D_{\text{f}} \right|$ are mostly independent of the sign of $g_{\text{bf}}$ in a wide region of the parameters. The symmetry for the boson-fermion attraction and repulsion is also shown in the instability condition of the HF ground state in Fig.~\ref{Fig-pd-rpa}.

In the interpretation of the forced oscillation explained above, the sign of the external force merely affects the phase of the oscillation. In addition, the effect of the sign of $g_{\text{bf}}$ on the initial state merely affects the phase of the initial position of the oscillator. As a result, the symmetry can appear in the results. This interpretation can also be confirmed by viewing of the time series of $d_{\text{f}}(t)$ before the Fourier transform. 

%
%
\section{Discussion \label{Sec-discussion}}
%
%
In this section, we discuss the results in Sec.~\ref{Sec-dynamic} in some theoretical points of view. In Sec.~\ref{Subsec-ES}, we discuss the effects of the quantum statistics and interactions on the dynamics by viewing of the excitation energy spectra. In Sec.~\ref{SubSec-HC}, we also discuss the correlation effects beyond the HF approximation. 

\subsection{Excitation energy spectra \label{Subsec-ES}}
According to Eqs.~(\ref{Eq-D-17}) and (\ref{Eq-D-22}), the excitation energy spectra, $E_{n} - E_{0} \equiv \Omega_{n}$ obtained in Eq.~(\ref{Eq-D-15}), are directly linked to the dynamics. 

In particular, the small amplitude excitations from the ground state are mostly construct from the $n = 0$ or $m = 0$ parts in Eqs.~(\ref{Eq-D-17}) and (\ref{Eq-D-22}), where these parts have the excitation energies $\Omega_{m}$ or $\Omega_{n}$, respectively. That is because the absolute values of the coefficients $C_{\alpha 0 m}$ and $C_{\alpha n 0}$ in Eq.~(\ref{Eq-D-18}) are much greater than those of $C_{\alpha n m}$ for $n \ne 0$ and $m \ne 0$ in the small amplitude excitations, i.e., $\big| \text{g} \big> \sim \big| 0 \big>$. 

In the present scheme, the amplitudes of the oscillations are determined by the interactions, $g_{\text{bb}}$ and $g_{\text{bf}}$, and deformation potentials, $V_{\text{b}}$ and $V_{\text{f}}$, in the initial state. 

In particular, for the small deformation potentials demonstrated in the previous section, the amplitude for the fermions is small, and that for bosons is chiefly determined by the boson-boson interaction, $g_{\text{bb}}$, because the boson-fermion interaction has small contributions to the deformation owing to the Fermi pressure as explained in Sec.~\ref{Sec-static}. In a rough estimation, the oscillations for bosons can be treated as the small amplitude excitations when $g_{\text{bb}} > g_{\text{bb}}^{(\text{c})}$ corresponding to the stability of the HF ground state in Eq.~(\ref{Sec-static-B-1}).

Here we first discuss the $g_{\text{bf}} = 0$ case, and then comment on the $g_{\text{bf}} \ne 0$ case. 

%
\begin{figure}[ht]
\begin{center}
\includegraphics[width=8.0cm]{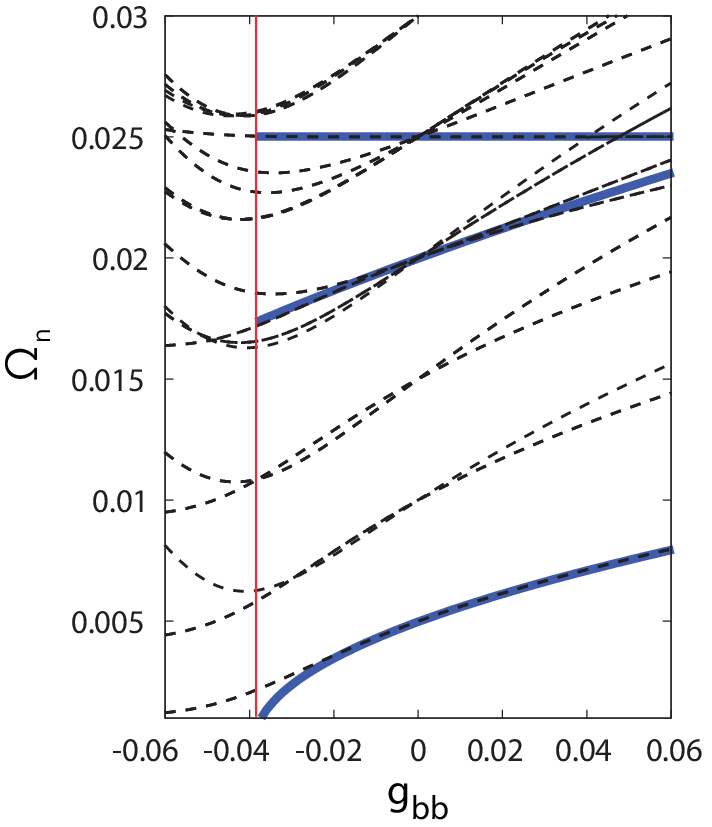}
\end{center}
\caption{(Color online) The excitation energy spectra, $\Omega_{n}$, when $g_{\text{bf}} = 0$. The dashed and solid lines indicate those in the exact diagonalization method and RPA, respectively. The left boundary of the spectra in RPA denoted by the thin-solid separator line indicates the instability boundary of the HF ground state.}
\label{F6}
\end{figure}
%

In Fig.~\ref{F6}, we show the excitation energy spectra, $\Omega_{n}$, in the $g_{\text{bf}} = 0$ case, where the bosons and fermions are independently moving as shown in Figs.~\ref{F3}, \ref{F2}(a), and \ref{F2}(b), and the spectra (denoted by the dashed line) are composed by the bosonic and fermionic spectra independent of each other. In fact, we can find the lowest fermionic spectra independent of $g_{\text{bb}}$ with $\Omega = 0.025$ in Fig.~\ref{F6}, and the other spectra below the lowest fermionic spectra are the bosonic spectra. Note that each of the spectra is degenerate, at least, double owing to the parity symmetry.

Here it should be noted that the quantum statistics directly affects the excitation energy spectra linked to the dynamics.
The results in RPA (denoted by the solid line) are also plotted in Fig.~\ref{F6}. The left boundary of the spectra in RPA (denoted by the thin-solid separator line) indicates the instability boundary of the HF ground state, i.e., $g_{\text{bb}} = g_{\text{bb}}^{(\text{c})}$. Here the results in RPA reproduce the exact ones for the lowest fermionic spectra exactly and for the lowest bosonic spectra except for the neighborhood of the instability boundary. Note that, in the particle-hole RPA, we can not obtain the multi-particle-multi-hole spectra, as shown in the bosonic spectra in Fig.~\ref{F6}, because of the ansatz for the $Q$ operators in Eq.~(\ref{Eq-rpa-5}). 

In the small amplitude excitations, the dominant peaks appear at the lowest spectra as explained above. It is confirmed by the results in Figs.~\ref{F2}(a), \ref{F2}(b), \ref{F3}(a), and \ref{F3}(b).

Moreover we can find out that, in the non-small amplitude excitations in Figs.~\ref{F3}(c) and \ref{F3}(d), the major peaks also appear at the lowest spectra, and the minor peaks correspond to the mixing of the lowest and next-lowest spectra. In addition, we also see that the invalidity of the HF approximation for the dynamics is directly linked to the instability of the HF ground state as shown in the lowest spectra in Fig.~\ref{F6}.

Lastly we comment on the $g_{\text{bf}} \ne 0$ case. 

The boson-fermion interaction slightly mixes the excitation energy spectra in Fig.~\ref{F6}. In particular, the mixing of the lowest bosonic and fermionic spectra is perturbative because of the energy difference of the lowest bosonic and fermionic spectra derived from the quantum statistics. As a result, this perturbation produces the forced oscillation as shown in Figs.~\ref{F2}(c), \ref{F2}(d), \ref{F4}, and \ref{F5}.

\subsection{Correlation effects \label{SubSec-HC}}
Here we discuss the correlation effects on the dynamics. We first introduce indicator parameters for the many-body correlations and the HF condition used in many-body physics, and then discuss the correlation effects by viewing of the calculational results of the indicator parameters. 
\subsubsection{The Hartree-Fock condition \label{SubSec-HC-HFC}}
The indicator parameters are defined as 
\begin{equation}
\sigma_{\text{b}} 
\equiv \frac{\text{tr}{\left[ \rho_{\text{b}}^{2} \right]}}{N_{\text{b}}^{2}} 
\label{Eq-HC-HFC-1}
\end{equation}
for bosons and 
\begin{equation}
\sigma_{\text{f}} 
\equiv \frac{\text{tr}{\left[ \rho_{\text{f}}^{2} \right]}}{N_{\text{f}}} 
\label{Eq-HC-HFC-2}
\end{equation}
for fermions with the one-body density matrices 
\begin{equation}
\left( \rho_{\text{b}} \right)_{\kappa \kappa^{\prime}} 
\equiv \big< \text{g} \big| b_{\kappa}^{\dagger} b_{\kappa^{\prime}} \big| \text{g} \big> 
\label{Eq-HC-HFC-3}
\end{equation}
and 
\begin{equation}
\left( \rho_{\text{f}} \right)_{\kappa \kappa^{\prime}} 
\equiv \big< \text{g} \big| c_{\kappa}^{\dagger} c_{\kappa^{\prime}} \big| \text{g} \big> 
\label{Eq-HC-HFC-4}
\end{equation}
for the ground state $\big| \text{g} \big>$. Here the indicator parameters are independent of selection of the base set for the density matrices because the traces in Eqs.~(\ref{Eq-HC-HFC-1}) and (\ref{Eq-HC-HFC-2}) are independent of it. In addition, $0 \le \sigma_{\text{b}} \le 1$ and $0 \le \sigma_{\text{f}} \le 1$ because of the definitions in Eqs.~(\ref{Eq-HC-HFC-1}) and (\ref{Eq-HC-HFC-2}) and positive-definite density matrices. 

In the HF approximation, we can obtain $\sigma_{\text{b}} = \sigma_{\text{f}} = 1$, which is called the HF condition. That is because the eigen-values of the density matrices agree with the occupation numbers of the single particles in the HF approximation. 

In general, the values of $\sigma_{\text{b}}$ and $\sigma_{\text{f}}$ are less than unity owing to the many-body correlations beyond the HF approximation. Thus they are usually used for the indicators of the correlations in many-body physics.

\subsubsection{Results \label{SubSec-GS}}
\begin{figure}[ht]
\begin{center}
\includegraphics[width=8.0cm]{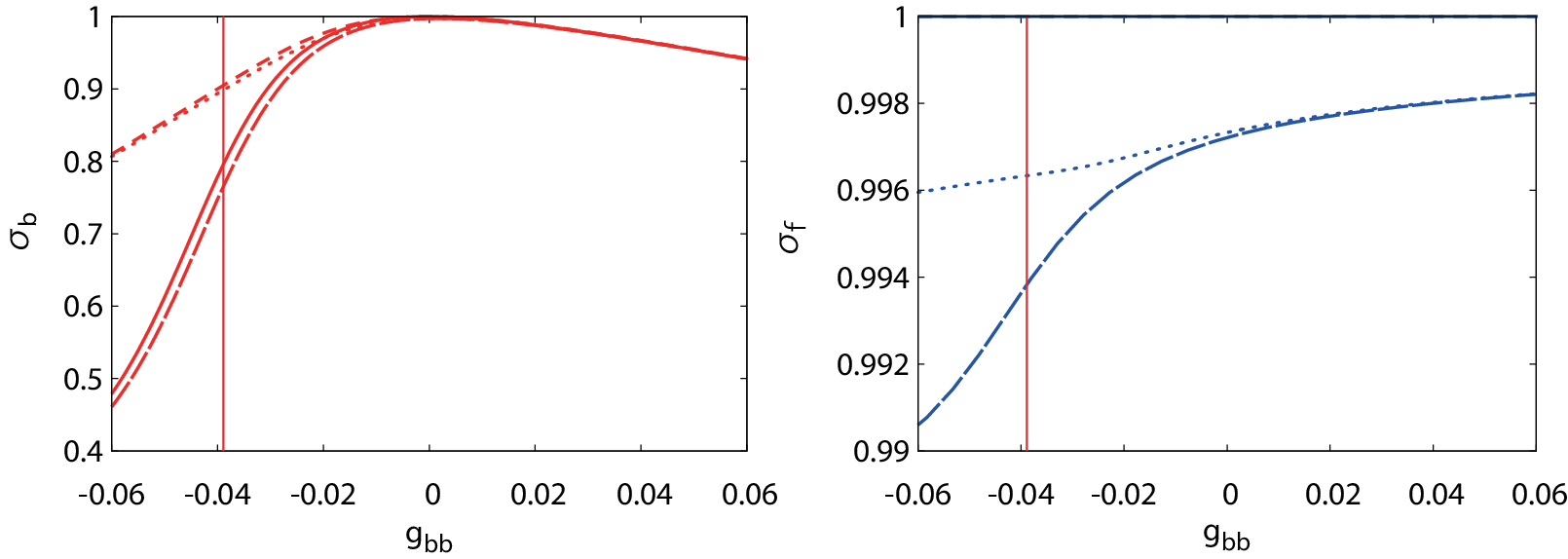}
\end{center}
\caption{(Color online) The indicator parameters $\sigma_{\text{b}}$ (a) and $\sigma_{\text{f}}$ (b). The solid and long-dashed lines indicate the results when $V_{\text{b}} = V_{\text{f}} = 0$ and $g_{\text{bf}} = 0$ and $- \big| g_{\text{bb}}^{(\text{c})} \big|$, respectively. The dashed and dotted lines indicate the results when $V_{\text{b}} = V_{\text{f}} = - 0.001$ and $g_{\text{bf}} = 0$ and $- \big| g_{\text{bb}}^{(\text{c})} \big|$, respectively. The thin-solid separator lines indicate $g_{\text{bb}} = g_{\text{bb}}^{(\text{c})}$.}
\label{F7}
\end{figure}

In Fig.~\ref{F7}, we show $\sigma_{\text{b}}$ and $\sigma_{\text{f}}$. 

%

First we focus on the results when $V_{\text{b}} = V_{\text{f}} = 0$ and $g_{\text{bf}} = 0$ (denoted by the solid lines). In this case, $\sigma_{\text{f}} = 1$ owing to the non-interacting fermions. On the other hand, $\sigma_{\text{b}}$ decreases from unity as $\big| g_{\text{bb}} \big|$ increases. In particular, $\sigma_{\text{b}}$ noticeably decreases around $g_{\text{bb}} = g_{\text{bb}}^{(\text{c})}$ (denoted by the thin-solid separator lines). 

Second we focus on the results when $g_{\text{bf}} = - \big| g_{\text{bb}}^{(\text{c})} \big|$ (denoted by the long-dashed and dotted lines). The boson-fermion interaction slightly decrease the values of $\sigma_{\text{b}}$ and $\sigma_{\text{f}}$. As shown in Fig.~\ref{F7}, the correlation effects of the boson-fermion interaction are much smaller than those of the boson-boson interaction. 

Third we finally focus on the results when $V_{\text{b}} = V_{\text{f}} = -0.001$ (denoted by the dashed and dotted lines). The small deformation potentials greatly increase the values of $\sigma_{\text{b}}$ and $\sigma_{\text{f}}$ around the instability boundary. 

In general, the many-body correlations increase as $\big| g_{\text{bb}} \big|$ and $\big| g_{\text{bf}} \big|$ increase as shown in Fig.~\ref{F7}; However, their effects on the dynamics are visually small, i.e., the HF approximation is valid, except for the instability region as demonstrated in Sec.~\ref{SubSec-dynamics}. 

In the instability region, the many-body correlations must be related to the rotational symmetry. That is because, as shown in Fig.~\ref{F7}, the small deformation potentials greatly reduce the correlations by breaking the rotational symmetry in the instability region; In the other region, the deformation potentials hardly affect the values of $\sigma_{\text{b}}$ and $\sigma_{\text{f}}$.

\subsubsection{Discussion \label{SubSubSec-GS}}

Here we discuss three points about the effects of the many-body correlations on the dynamics. 

First we discuss a relationship between the many-body correlations and rotational symmetry in the instability region in terms of the HF theory. The relationship is demonstrated in Fig.~\ref{F7} as explained above. 

In the HF theory, the instability induces the deformed HF ground state with an infinitesimal deformation potential called a seed of the deformation. It is consistent with the great reduction of the many-body correlations with the small deformation potentials in the instability region in Fig.~\ref{F7}. 

The deformed HF ground state can not directly be used for an approximation of the exact ground state. That is because the deformation direction of the deformed state is determined by the artificial seed, and the different seed produces the different state. Furthermore all of the produced states are degenerate because of the rotational symmetry. In principle, the ground state must have the rotational symmetry in the finite quantum systems. 

As an extended HF theory, we can consider a superposition state of all of the deformed HF ground states and determine the coefficients by estimation of the full energy beyond the mean-field energy. This scheme is called the generator coordinate method or projection method and studied well in nuclear physics for deformed nuclei~\cite{RS}. 

In the projection method, we can obtain the approximate states for the ground state and some low-energy excited states, and the rotational symmetry of the states is restored. 

In this sense, the many-body correlations in the instability region can be interpreted as the superposition of the deformed HF ground state for the restoration of the broken symmetry. In addition, the bosonic low-frequency modes in the dynamics in Sec.~\ref{SubSec-dynamics} can also be interpreted as the restoration modes of the broken symmetry in terms of the projection method. 

Second we discuss another relationship between the many-body correlations and macroscopic behaviors of the toroidal gases in the thermodynamical limit, where the macroscopic behaviors can be described in quantum field theory except for the finite-size effects. 

According to quantum field theory, the ground state can exhibit the spontaneous symmetry breaking with the inequivalent vacua and Nambu-Goldstone mode in the thermodynamical limit~\cite{AFT}, where quantum field theory must be exact, unlike the HF theory. 

The symmetry broken states exhibit macroscopic objects and can not transfer to each other by any quantum-mechanical operators. In fact, the dynamics in Sec.~\ref{SubSec-dynamics} must get close to the macroscopic behaviors, where the dominant oscillation mode is almost at $\omega = 0$ owing to the correlation effects and indicates the immobile object in a finite time~\footnote{%
Although the HF theory must be different from quantum field theory, the macroscopic behaviors may be analogous to the soliton-like behaviors in the HF approximation. 
}. In other words, the dynamics must be dominated by the finite-size effect. 

In terms of quantum field theory, the many-body correlations related to the symmetry can be interpreted as a precursor of the spontaneous symmetry breaking, or the phase transition, in the finite quantum systems. 

Third we discuss a mechanism of the many-body correlations in quantum mechanics for the finite systems. 

In quantum mechanics, the many-body correlations related to the symmetry must be due to existence of quasi-degenerate states. That is because the perturbative deformation potentials greatly affect the correlations as shown in Fig.~\ref{F7}, and the perturbation can not induce such a great effect on the states except for the quasi-degenerate states according to the perturbation theory. 

Since a degeneracy in quantum mechanics must be based on a symmetry and its irreducible orthogonal states, the quasi-degeneracy must also be based on the rotational symmetry and approximately-orthogonal states, $\big< \text{g}(\theta_{\text{d}}) \big| \big| \text{g}({\theta_{\text{d}}}^{\prime}) \big> \approx \delta_{\theta_{\text{d}}, {\theta_{\text{d}}}^{\prime}}$, where we introduce the ground state, $\big| \text{g}(\theta_{\text{d}}) \big>$ for $H + H^{\prime}(\theta_{\text{d}})$ with the  deformation direction $\theta_{\text{d}}$ ($= 0$ for the initial state in the dynamics). The approximate orthogonality can clearly appear near the the thermodynamical limit corresponding to the phase transition in quantum field theory and also gradually appear in the small-number systems. 

In the finite quantum many-body physics, the quasi-degeneracy must be important for the correlation effects. 

Although the above three points are based on the different theories in mathematics, the findings must be consistent in terms of physics. Thus the studies on the dynamics in the cold atoms must contribute to those on the theories. 

\section{Summary and perspective \label{Sec-summary}}

In the present paper, we study the dynamics of the quasi-one-dimensional boson-fermion mixtures of the toroidal gases in the exact diagonalization method. Especially we focus on the effects of the quantum statistics and many-body correlations. 

In Sec.~\ref{Sec-form}, we formulate the model and dynamics, and also explain the calculational schemes.  

In Sec.~\ref{Sec-static}, we explain the static properties of the toroidal gases as a background knowledge. 

In Sec.~\ref{Sec-dynamic}, we demonstrate the dynamics in the symmetric situations for the bosons and fermions. In particular, we show the quite different behaviors of the bosons and fermions owing to the quantum statistics, and also demonstrate the weak correlation effects in the stability region and strong correlation effects with the low-frequency modes in the instability region. In addition, we reveal the strong influence of the bosons on the fermions as a forced oscillator. 

In Sec.~\ref{Sec-discussion}, we discuss the effects of the quantum statistics and many-body correlations on the dynamics by viewing of the excitation energy spectra and indicator parameters for the correlations. 

The present study must be linked to actual experimental situations; However, although the toroidal bosons have been realized and studied well, the experimental studies on the boson-fermion mixtures of the toroidal gases have not been demonstrated yet. We hope realization of the studies in future experiments.

Our scheme in the present paper must be applicable to various studies on the dynamics and many-body correlations in the cold atoms. In particular, studies on the many-body correlations may be important in the recently-realized systems with the dipolar interaction~\cite{dipole-dipole} and spin-orbit interaction~\cite{spin-orbit} because these systems may have rich phase structures related to the correlations. 

Lastly we comment on the previous studies on the collective excitations of the boson-fermion mixtures~\cite{sogo, mr}. These studies cover the three-dimensional gases and are based on the HF approximation. Although the present study covers the quasi-one-dimensional gases and is beyond the HF approximation, the results in these studies show some analogous behaviors, e.g., the different behaviors of the bosons and fermions, the influence of the bosons on the fermions as a forced oscillator, and so on. These behaviors may be independent of the dimensionalities and many-body correlations.  

In future works, we should study the asymmetric situations for the bosons and fermions, which are not demonstrated in the present paper and must be treated in the same formulation. In addition, we should also study the toroidal gases with the very strong interactions, which are not treated in the present paper.

\acknowledgments
The authors thank Toru Suzuki and Tomoyuki Maruyama for useful discussions and comments. One of the authors (T. N.) also thanks Rina Kanamoto. This work was supported by the MEXT program "Support Program for Improving Graduate School Education" and KAKENHI (22540414). 

\appendix
\section{Random phase approximation \label{APP_2}}

In RPA~\cite{RS}, we approximately obtain a part of the excitation energies 
\begin{equation}
\Omega_{n} 
\equiv E_{n} - E_{0} 
\label{Eq-rpa-1}
\end{equation}
and $Q$ operators 
\begin{equation}
Q_{n} 
\equiv \big| 0 \big> \big< n \big| 
\label{Eq-rpa-2}
\end{equation}
for the quantum number $n$ ($\ge 1$) in Eqs.~(\ref{Eq-D-15}) and (\ref{Eq-D-16}) by using the variational method in a variational ansatz for the $Q$ operators to minimize the excitation energies. 

Because of the orthonormal relations, the $Q$ operators must satisfy 
\begin{equation}
\big< 0 \big| Q_{n} Q_{n^{\prime}}^{\dagger} \big| 0 \big> 
= \big< 0 \big| \big[ Q_{n}, Q_{n^{\prime}}^{\dagger} \big]_{-} \big| 0 \big> 
= \delta_{n, n^{\prime}}, 
\label{Eq-rpa-3}
\end{equation}
where the commutation relation in the middle part is a convention. Note that the $Q$ operators are not the boson operators owing to the definition in Eq.~(\ref{Eq-rpa-2}). 

The $Q$ operators determine the excitation energies in Eq.~(\ref{Eq-rpa-1}) as 
\begin{equation}
\Omega_{n} 
= \frac{\big< 0 \big| Q_{n} \big[ H, Q_{n}^{\dagger} \big]_{-} \big| 0 \big>}{\big< 0 \big| Q_{n} Q_{n}^{\dagger} \big| 0 \big>} 
= \frac{\big< 0 \big| \big[ Q_{n}, \big[ H, Q_{n}^{\dagger} \big]_{-} \big]_{-} \big| 0 \big>}{\big< 0 \big| \big[ Q_{n}, Q_{n}^{\dagger} \big]_{-} \big| 0 \big>}, 
\label{Eq-rpa-4}
\end{equation}
where the right part is also a convention. 

In the particle-hole RPA, we assume the $Q$ operators as 
\begin{eqnarray}
Q_{n}^{\dagger} 
&\approx& \sum_{k \ne 0} \frac{1}{\sqrt{N_{\text{b}}}} \left( X_{n}^{k} b_{k}^{\dagger} b_{0} - Y_{n}^{k} b_{0}^{\dagger} b_{k} \right) 
\nonumber \\ &&
+ \sum_{\left| p \right| \le \kappa_{\text{f}}} \sum_{\left| h \right| > \kappa_{\text{f}}} \left( x_{n}^{p h} c_{p}^{\dagger} c_{h} - y_{n}^{p h} c_{h}^{\dagger} c_{p} \right) 
\label{Eq-rpa-5}
\end{eqnarray}
with the Fermi momentum $\kappa_{\text{f}}$ and variational parameters $X_{n}^{k}$, $Y_{n}^{k}$, $x_{n}^{p h}$, and $y_{n}^{p h}$. 

In addition, the ground state $\big| 0 \big>$ in Eqs.~(\ref{Eq-rpa-3}) and (\ref{Eq-rpa-4}) is approximated by the HF ground state $\big| 0_{\text{HF}} \big>$, where $b_{k} \big| 0_{\text{HF}} \big> = c_{p} \big| 0_{\text{HF}} \big> = c_{h}^{\dagger} \big| 0_{\text{HF}} \big> = 0$ and so on. Then the particle-hole RPA agrees with the time-dependent HF approximation in the small amplitude excitations from the HF ground state. That is because the HF state in Eq.~(\ref{Eq-calculation-HF-1}) in the small amplitude excitations can be constructed from the particle-hole fluctuations in Eq.~(\ref{Eq-rpa-5}). 

As a result, we obtain the variational equations, $\delta{\Omega_{n}} = 0$, denoted by 
\begin{equation}
A \vec{z}_{n} 
= \Omega_{n} \vec{z}_{n} 
\label{Eq-rpa-6}
\end{equation}
with 
\begin{equation}
\vec{z}_{n} 
\equiv \left[ 
\begin{array}{c}
\left\{ X_{n}^{k};~ k \ne 0 \right\} 
\\ 
\left\{ Y_{n}^{k};~ k \ne 0 \right\} 
\\ 
\left\{ x_{n}^{p h};~ \left| p \right| \le \kappa_{\text{f}},~ \left| h \right| > \kappa_{\text{f}} \right\} 
\\ 
\left\{ y_{n}^{p h};~ \left| p \right| \le \kappa_{\text{f}},~ \left| h \right| > \kappa_{\text{f}} \right\} 
\end{array}
\right] 
\label{Eq-rpa-7}
\end{equation}
and 
\begin{equation}
A 
\equiv \left[ 
\begin{array}{cccc}
\left\{ A_{1 1}^{k k^{\prime}} \right\} & 
\left\{ A_{1 2}^{k k^{\prime}} \right\} & 
\left\{ A_{1 3}^{k p^{\prime} h^{\prime}} \right\} & 
\left\{ A_{1 4}^{k p^{\prime} h^{\prime}} \right\} 
\\ 
\left\{ A_{2 1}^{k k^{\prime}} \right\} & 
\left\{ A_{2 2}^{k k^{\prime}} \right\} & 
\left\{ A_{2 3}^{k p^{\prime} h^{\prime}} \right\} & 
\left\{ A_{2 4}^{k p^{\prime} h^{\prime}} \right\} 
\\ 
\left\{ A_{3 1}^{p h k^{\prime}} \right\} & 
\left\{ A_{3 2}^{p h k^{\prime}} \right\} & 
\left\{ A_{3 3}^{p h p^{\prime} h^{\prime}} \right\} & 
\left\{ A_{3 4}^{p h p^{\prime} h^{\prime}} \right\} 
\\ 
\left\{ A_{4 1}^{p h k^{\prime}} \right\} & 
\left\{ A_{4 2}^{p h k^{\prime}} \right\} & 
\left\{ A_{4 3}^{p h p^{\prime} h^{\prime}} \right\} & 
\left\{ A_{4 4}^{p h p^{\prime} h^{\prime}} \right\} 
\end{array}
\right], 
\label{Eq-rpa-8}
\end{equation}
where the superscripts in Eq.~(\ref{Eq-rpa-8}) vary in the same ranges in Eq.~(\ref{Eq-rpa-7}) in the same order, and $k$, $p$, and $h$ indicate the column components; $k^{\prime}$, $p^{\prime}$, and $h^{\prime}$ indicate the row components. The Hermitian conjugate becomes 
\begin{equation}
\vec{z}_{n}^{\dagger} D A 
= \Omega_{n} \vec{z}_{n}^{\dagger} D 
\label{Eq-rpa-9}
\end{equation}
with 
\begin{equation}
D 
\equiv \left[ 
\begin{array}{cccc}
\left\{ \delta_{k, k^{\prime}} \right\} & 
\left\{ 0 \right\} & 
\left\{ 0 \right\} & 
\left\{ 0 \right\} 
\\ 
\left\{ 0 \right\} & 
\left\{ - \delta_{k, k^{\prime}} \right\} & 
\left\{ 0 \right\} & 
\left\{ 0 \right\} 
\\ 
\left\{ 0 \right\} & 
\left\{ 0 \right\} & 
\left\{ \delta_{p, p^{\prime}} \delta_{h, h^{\prime}} \right\} & 
\left\{ 0 \right\} 
\\ 
\left\{ 0 \right\} & 
\left\{ 0 \right\} & 
\left\{ 0 \right\} & 
\left\{ - \delta_{p, p^{\prime}} \delta_{h, h^{\prime}} \right\} 
\end{array}
\right] 
\label{Eq-rpa-10}
\end{equation}
because $A^{\dagger} = D A D$, $D^{2} = 1$, and $\Omega_{n}^{*} = \Omega_{n}$. 

The matrix elements in Eq.~(\ref{Eq-rpa-8}) are obtained as 
\begin{eqnarray}
&& 
A_{1 1}^{k k^{\prime}} 
= - A_{2 2}^{k k^{\prime}} 
= \left[ \frac{k^{2}}{2 m_{\text{b}} R^{2}} + g_{\text{bb}} \left( N_{\text{b}} - 1 \right) \right] \delta_{k, k^{\prime}}, 
\nonumber \\ && 
A_{1 2}^{k k^{\prime}} 
= - A_{2 1}^{k k^{\prime}} 
= g_{\text{bb}} \left( N_{\text{b}} - 1 \right) \delta_{0, (k + k^{\prime})}, 
\nonumber \\ && 
A_{1 3}^{k p^{\prime} h^{\prime}} 
= - A_{2 4}^{k p^{\prime} h^{\prime}} 
= g_{\text{bf}} \sqrt{N_{\text{b}}} \delta_{p^{\prime}, (h^{\prime} + k)}, 
\nonumber \\ && 
A_{1 4}^{k p^{\prime} h^{\prime}} 
= - A_{2 3}^{k p^{\prime} h^{\prime}} 
= g_{\text{bf}} \sqrt{N_{\text{b}}} \delta_{h^{\prime}, (p^{\prime} + k)}, 
\nonumber \\ && 
A_{3 1}^{p h k^{\prime}} 
= - A_{4 2}^{p h k^{\prime}} 
= g_{\text{bf}} \sqrt{N_{\text{b}}} \delta_{p, (h + k^{\prime})}, 
\nonumber \\ && 
A_{3 2}^{p h k^{\prime}} 
= - A_{4 1}^{p h k^{\prime}} 
= g_{\text{bf}} \sqrt{N_{\text{b}}} \delta_{h, (p + k^{\prime})}, 
\nonumber \\ && 
A_{3 3}^{p h p^{\prime} h^{\prime}} 
= - A_{4 4}^{p h p^{\prime} h^{\prime}} 
= \frac{p^{2} - h^{2}}{2 m_{\text{f}} R^{2}} \delta_{p, p^{\prime}} \delta_{h, h^{\prime}}, 
\nonumber \\ && 
A_{3 4}^{p h p^{\prime} h^{\prime}} 
= A_{4 3}^{p h p^{\prime} h^{\prime}} 
= 0. 
\label{Eq-rpa-11}
\end{eqnarray}

The orthonormal relations in Eq.~(\ref{Eq-rpa-3}) become 
\begin{equation}
\vec{z}_{n}^{\dagger} D \vec{z}_{n^{\prime}} 
= \delta_{n, n^{\prime}} 
\label{Eq-rpa-12}
\end{equation}
and can be kept in Eq.~(\ref{Eq-rpa-6}) except for degeneracy because of the real-valued energy $\Omega_{n}$ and Eq.~(\ref{Eq-rpa-9}). 

Eq.~(\ref{Eq-rpa-6}) has pairs of the symmetric solutions for the positive and negative energies (denoted by $\Omega_{n} = - \Omega_{- n} > 0$) owing to the time-reversal symmetry. Of course, the physical solutions must be the positive solutions (for $n \ge 1$) because of the definition in Eq.~(\ref{Eq-rpa-1}). 

Although the physical lowest eigen-value $\Omega_{1}$ in Eq.~(\ref{Eq-rpa-4}) must be a real positive number for the exact ground state $\big| 0 \big>$, it can go into a pure imaginary number for the HF ground state $\big| 0_{\text{HF}} \big>$. The imaginary solution indicates the instability of the HF ground state for the particle-hole fluctuations.


\end{document}